\documentclass[10pt,journal,compsoc]{IEEEtran}
\ifCLASSOPTIONcompsoc
  \usepackage[nocompress]{cite}
\else
  \usepackage{cite}
\fi
\pdfoutput=1
\usepackage{graphicx}
\usepackage{bm}
\usepackage[utf8]{inputenc}
\usepackage{graphicx}
\usepackage{epstopdf}
\usepackage{multirow}
\usepackage{amsmath}
\usepackage{tabularx}
\usepackage{booktabs}
\usepackage{float}
\usepackage[linesnumbered,algoruled,boxed,lined]{algorithm2e}
\usepackage{array}
\usepackage{color}
\usepackage{amssymb}
\usepackage{cite}
\usepackage{soul}

\newtheorem{proposition}{Proposition}
\usepackage{adjustbox}
\usepackage{subfigure}
\usepackage{setspace}
\usepackage{bm}
\usepackage{utfsym}

\ifCLASSINFOpdf
\else
\fi
\begin{document}
%
\title{Cooperative UAV-mounted RISs-assisted Energy-efficient Communications}
%
%
%
%

\author{Hongyang Pan, Yanheng Liu, Geng Sun,~\IEEEmembership{Senior Member,~IEEE}, Qingqing Wu,~\IEEEmembership{Senior Member,~IEEE}, \\Tierui Gong,~\IEEEmembership{Member,~IEEE}, Pengfei Wang,~\IEEEmembership{Member,~IEEE}, \\Dusit Niyato,~\IEEEmembership{Fellow,~IEEE}, and Chau Yuen,~\IEEEmembership{Fellow,~IEEE}
        \thanks{\quad This study is supported in part by the National Natural Science Foundation of China (62272194, 62471200), in part by the Science and Technology Development Plan Project of Jilin Province (20250101027JJ), and in part by the National Research Foundation, Singapore and Infocomm Media Development Authority under its Future Communications Research \& Development Programme (FCP-NTU-RG-2024-025). (\textit{Corresponding author: Geng Sun}).}

        \thanks{\quad Hongyang Pan is with the College of Computer Science and Technology, Jilin University, Changchun 130012, China, and also with the Information Science and Technology College, Dalian Maritime University, Dalian 116026, China (E-mail: panhongyang18@foxmail.com).}

        \thanks{\quad Yanheng Liu is with the College of Computer Science and Technology, Jilin University, Changchun 130012, China, and also with the Key Laboratory of Symbolic Computation and Knowledge Engineering of Ministry of Education, Jilin University, Changchun 130012, China (E-mail: yhliu@jlu.edu.cn).}

        \thanks{\quad Geng Sun is with the College of Computer Science and Technology, Jilin University, Changchun 130012, China, and with Key Laboratory of Symbolic Computation and Knowledge Engineering of Ministry of Education, Jilin University, Changchun 130012, China; he is also affiliated with the College of Computing and Data Science, Nanyang Technological University, Singapore 639798 (E-mail: sungeng@jlu.edu.cn).}

        \thanks{\quad Qingqing Wu is with the Department of Electronic Engineering, Shanghai Jiao Tong University, Shanghai, China (E-mail: qingqingwu@sjtu.edu.cn.).}

        \thanks{\quad Tierui Gong and Chau Yuen are with the School of Electrical and Electronics Engineering, Nanyang Technological University, Singapore 639798 (E-mail: trgTerry1113@gmail.com, chau.yuen@ntu.edu.sg).}

        \thanks{\quad Pengfei Wang is with the School of Computer Science and Technology, Dalian University of Technology, Dalian 116024, China (E-mail: wangpf@dlut.edu.cn).}

        \thanks{\quad Dusit Niyato is with the School of Computer Science and Engineering, Nanyang Technological University, Singapore 639798 (E-mail: dniyato@ntu.edu.sg.).}
        
        \thanks{\quad This manuscript has been accepted by IEEE Transactions on Mobile Computing, DOI: 10.1109/TMC.2025.3579597.}
}

%
%

\markboth{Journal of \LaTeX\ Class Files,~Vol.~14, No.~8, August~2015}%
{Shell \MakeLowercase{\textit{et al.}}: Bare Demo of IEEEtran.cls for Computer Society Journals}
%



\IEEEtitleabstractindextext{%
\begin{abstract}
Cooperative reconfigurable intelligent surfaces (RISs) are promising technologies for 6G networks to support a great number of users. Compared with the fixed RISs, the properly deployed RISs may improve the communication performance with less communication energy consumption, thereby improving the energy efficiency. In this paper, we consider a cooperative unmanned aerial vehicle-mounted RISs (UAV-RISs)-assisted cellular network, where multiple RISs are carried and enhanced by UAVs to serve multiple ground users (GUs) simultaneously such that achieving the three-dimensional (3D) mobility and opportunistic deployment. Specifically, we formulate an energy-efficient communication problem based on multi-objective optimization framework (EEComm-MOF) to jointly consider the beamforming vector of base station (BS), the location deployment and the discrete phase shifts of UAV-RIS system so as to simultaneously maximize the minimum available rate over all GUs, maximize the total available rate of all GUs, and minimize the total energy consumption of the system, while the transmit power constraint of BS is considered. To comprehensively solve EEComm-MOF which is an NP-hard and non-convex problem with constraints, a non-dominated sorting genetic algorithm-II with a continuous solution processing mechanism, a discrete solution processing mechanism, and a complex solution processing mechanism (INSGA-II-CDC) is proposed. Simulations results demonstrate that the proposed INSGA-II-CDC can solve EEComm-MOF effectively and outperforms other benchmarks under different parameter settings. Moreover, the stability of INSGA-II-CDC and the effectiveness of the improved mechanisms are verified. Finally, the implementability analysis of the algorithm is given.
\end{abstract}
}

\maketitle

\IEEEdisplaynontitleabstractindextext

%
\IEEEpeerreviewmaketitle
\IEEEraisesectionheading{\section{Introduction}\label{sec:introduction}}
\IEEEPARstart{6}{G} wireless networks are facing the challenge of accommodating a large number of users while meeting their ever-increasing demands for spectral and energy efficiency \cite{DBLP:journals/wc/HuangHAZYZRD20}, \cite{9952197}. To address this issue, a revolutionary technology called reconfigurable intelligent surface (RIS) has emerged as a feasible solution for intelligently reconfiguring the wireless propagation environment to enable energy-efficient communication \cite{10109654}, \cite{DBLP:journals/twc/WuZ19}, \cite{DBLP:journals/twc/ChenWCNH23}. An RIS can be described as a meta-surface comprising numerous passive and cost-effective elements, where each element of the RIS can independently adjust its phase shifts of the impinging radio waves, thereby transforming the wireless environment from a highly unpredictable space to a partially deterministic one \cite{10158690}, \cite{10088448}. In a cellular network equipped with an RIS, a base station (BS) transmits signals to the RIS controller, optimizing the characteristics of the incident waves. Consequently, the RIS acts as a reflector, enhancing service quality of users. When RIS is deployed properly, it has the potential to be more energy-efficient than the amplify-and-forward technology and backscatter technology \cite{10050148}. Furthermore, compared with continuous phase shifts, it is practical to consider the discrete phase shifts of an RIS \cite{DBLP:journals/wcl/GaoCHY21}.
\par To establish a virtual line-of-sight (LoS) communication environment, RISs are commonly deployed in fixed locations, such as facades \cite{10109654, DBLP:journals/twc/NingHWWYYZ24, DBLP:journals/jsac/CaoYHYRNH21}. However, the location of RIS deployment can influence energy efficiency and service quality. \textcolor{black}{In this regard, using mobile vehicles such as unmanned aerial vehicles (UAVs) and balloons to deploy RISs is promising \cite{DBLP:conf/globecom/ZhangSB19, DBLP:journals/jsac/LiuLC21, DBLP:journals/twc/KhaliliMZJYJ22}. Compared to balloons-mounted RISs, UAV-mounted RISs (UAV-RISs) are more practical for flexible and opportunistic deployment to enhance the communication performance \cite{DBLP:journals/twc/ChengPHAYD22, DBLP:journals/jsac/WuXZNASS21a}, since UAVs generally have a faster speed than balloons. Moreover, using UAVs means more stable communication links, as balloons are vulnerable by the external disturbances like wind, causing them to shift from their intended locations and angles \cite{10793319}, while UAVs can use advanced control systems to counteract them. Although using tethered balloons can enhance communication stability \cite{DBLP:journals/icl/SudheeshMMSM18}, the deployment freedom of RISs will be correspondingly damaged. Moreover, due to the low costs of UAVs and RISs, UAV-RIS has good scalability. For example, when the ground users (GUs) in the network suffer from reduced signal reception due to changes in channel conditions, increasing the number of UAV-RISs can enhance the strength of the reflected signals.} However, it is important to consider that the use of UAVs introduces additional energy consumption, which must also be taken into account during system design and operation.

\par With the increasing number of GUs, using a single RIS of limited size to simultaneously serve all GUs becomes challenging \cite{DBLP:journals/twc/YangCSXSPC22}. Moreover, expanding the reflection area of RIS is also not reasonable, since UAVs with limited hardware resources are hard to load a large enough size RIS, so as to satisfy the communication requirements \cite{DBLP:conf/iecon/ZhouQ018}. To this end, the concept of cooperative RISs has emerged as a promising area to enhance system capacity \cite{10025789}. By incorporating UAVs into the system, each UAV-RIS can be deployed properly, thereby boosting the received signal strength. However, there is mutual interference among different links, imposing higher requirements on the location deployment of UAV-RISs.

\par This work considers a ground BS and several GUs, while the direct links from the BS to GUs are entirely unavailable. Thus, several cooperative UAV-RISs are employed in a three-dimensional (3D) space to serve all GUs simultaneously. \textcolor{black}{Different from the previous works that focused on a single UAV-RIS to serve the GUs \cite{DBLP:journals/tits/ZhaoSNXGZ24}, \cite{DBLP:journals/wcl/MohamedHH22}, or investigated two-dimensional (2D) mobility of UAV-RISs \cite{DBLP:conf/wcnc/WuGLGKK24}, \cite{DBLP:conf/globecom/GeZW21}, the considered scenario of this work is more intricate. Moreover, the proposed framework can be extended to some scenarios, such as mountain disaster rescue scenario and urban communication scenario. In these cases, UAV-RISs-assisted cellular network can efficiently utilize the 3D mobility and the opportunistic deployment to achieve low-delay communications, which can further improve rescue efficiency and user experience, respectively.} The primary contributions of this paper are as follows: 
\color{black}
\begin{itemize}
	\item \textbf{\textit{Cooperative UAV-RISs-assisted Energy-efficient Communication System:}} We consider a cooperative UAV-RISs-assisted cellular network, where UAVs mount and enhance multiple RISs to simultaneously serve a number of GUs. This setup enables 3D mobility and allows for opportunistic deployment of RISs, providing a flexible and adaptive coverage. Such a communication system is widely used and can be extended to a range of practical scenarios, such as mountain disaster rescue communications and urban communications. These practical scenarios are closely aligned with real-world requirements, where traditional communication links may be unavailable or unreliable, and the system can effectively address the challenges posed by unavailable links.
	
	\item \textbf{\textit{Energy-efficient Communication Problem based on Multi-objective Optimization Framework Formulation:}} In the considered system, three optimization objectives are maximizing the minimum available rate over all GUs, maximizing the total available rate of all GUs, and minimizing the total energy consumption of the system, which correspond to fair service, system capacity, and system cost, respectively, while there are trade-offs among them. Thus, an energy-efficient communication problem based on multi-objective optimization framework (EEComm-MOF) is formulated. To the best of our knowledge, it is the first work to jointly take into account deploying multiple UAV-RISs to serve multiple GUs by adjusting 3D locations of UAV-RISs, discrete phase shifts of UAV-RISs and beamforming vector of BS, while simultaneously considering different optimization objectives, making the problem non-trivial.
	
	\item \textbf{\textit{Multi-objective Optimization with Improved Non-dominated Sorting Genetic Algorithm-II:}} Given the NP-hardness and non-convexity of EEComm-MOF, we propose an improved non-dominated sorting genetic algorithm-II with a continuous solution processing mechanism, a discrete solution processing mechanism, and a complex solution processing mechanism (INSGA-II-CDC) to solve the problem. Specifically, the continuous solution processing mechanism can exploit the better deployed locations of UAV-RISs, so as to improve the convergence rate. In addition, the discrete solution processing mechanism as well as the complex solution processing mechanism enable the algorithm to handle discrete phase shifts of UAV-RISs and beamforming vector of BS, respectively, thereby achieving a better solution set distribution. Thus, these three improved mechanisms can jointly enhance the search capability of the algorithm in the limited iterations.
	
	\item \textbf{\textit{Performance Evaluations and Analyses:}} \textcolor{black}{Through simulations, we evaluate the performance of the proposed INSGA-II-CDC for solving the formulated EEComm-MOF under different settings. \textcolor{black}{Specifically, the performance of the INSGA-II-CDC in terms of the convergence and optimality, stability, effectiveness of improved mechanisms, and CPU running time is verified.} For a cellular network with $5$ GUs, compared to the suboptimal value obtained by other benchmarks on the corresponding objective, the minimum available rate of the proposed approach can be enhanced by $74.62\%$, and the total available rate can be improved by $64.45\%$, while the corresponding energy consumption is saved by $10.55\%$. Similarly, compared to the suboptimal value obtained by other benchmarks on the corresponding objective for a cellular network with $10$ GUs, the minimum available rate, the total available rate, and the energy consumption of the proposed approach can be increased by $43.75\%$, increased by $89.57\%$, and reduced by $13.60\%$, respectively. Finally, the implementability analysis of the algorithm is given.}
\end{itemize}
\color{black}

\par The remainder of the paper is organized as follows. The related work is reviewed in Section \ref{sec:Related Work}. Section \ref{sec:Models} presents the system model. The EEComm-MOF is formulated in Section \ref{sec:Problem Formulation and Analysis}. Section \ref{sec:Algorithm} gives the algorithm for EEComm-MOF. Section \ref{sec:Simulation Results} provides the simulation results and Section \ref{sec:Conclusion} concludes the paper.

\par \emph{Notations}: The space of $M \times N$ complex-valued matrices is denoted by $\mathbb{C}^{M \times N}$. For a vector $\mathbf{x}$, its Euclidean norm is represented by $\left\|\mathbf{x}\right\|$. The function $\operatorname{diag}(\mathbf{x})$ represents a diagonal matrix whose diagonal elements are given by the entries of $\mathbf{x}$. The imaginary unit of a complex number is denoted by $j = \sqrt{-1}$. The symbol $\mathbf{x}^T$ represents the transpose of the vector $\mathbf{x}$, while the symbol $\mathbf{x}^H$ represents the conjugate transpose of the vector $\mathbf{x}$. Finally, the Kronecker product operation is denoted as $\otimes$.

\begin{table*}[htb]
	\begin{center}
		\scriptsize
		\caption{\textcolor{black}{Main contributions of related works}}
 		\setlength{\tabcolsep}{0.6mm}
		\label{Reference1}
		{\begin{tabular}{|c|c|c|c|c|c|c|c|c|c|c|c|}\cline{1-11}
           & \multicolumn{2}{c|}{Scenario complexity}& \multicolumn{3}{c|}{Decision variable}  &  \multicolumn{4}{c|}{Optimization objective}& Method\\
                \cline{1-11}
				Reference  & \begin{tabular}{c} 
		Multiple \\RISs\end{tabular}  &\begin{tabular}{c} Multiple \\GUs\end{tabular} &\begin{tabular}{c} 3D location\\ deployment\\ of RIS  \end{tabular} &\begin{tabular}{c}Discrete\\phase shifts\\ of RIS \end{tabular} & \begin{tabular}{c} Beamforming\\ vector of BS \end{tabular}    &\begin{tabular}{c} Minimum \\available\\rate\end{tabular} &\begin{tabular}{c} Total\\available\\rate\end{tabular} &\begin{tabular}{c} UAV flight\\ energy \\ consumption \end{tabular} &\begin{tabular}{c} Communication\\ energy \\ consumption \end{tabular}&\begin{tabular}{c} Multi-objective \\evolutionary \\algorithm\end{tabular} \\ 
				\cline{1-11}
			
				
			\textbf{\cite{10109654}}  	 
				&$\usym{2715}$ &$\checkmark$ 
				&$\usym{2715}$ 
				&$\usym{2715}$
				&$\checkmark$
                &$\usym{2715}$
                &$\usym{2715}$
                &$\usym{2715}$
                &$\checkmark$
                &$\usym{2715}$
                \\\cline{1-11}

                \textbf{\cite{10025789}} &$\checkmark$  &$\usym{2715}$  &$\usym{2715}$ &$\usym{2715}$ 
				&$\usym{2715}$
                &$\usym{2715}$
                &$\checkmark$
                &$\usym{2715}$
                &$\usym{2715}$
                &$\usym{2715}$
				\\\cline{1-11}

				\textbf{\cite{DBLP:journals/tits/ZhaoSNXGZ24}}	 
				&$\usym{2715}$
				&$\checkmark$
				&$\usym{2715}$
				&$\usym{2715}$
				&$\usym{2715}$
				&$\usym{2715}$
				&$\checkmark$
				&$\usym{2715}$
				&$\usym{2715}$
				&$\usym{2715}$
				\\\cline{1-11}

                \textbf{\cite{DBLP:journals/wcl/GaoCHY21}} &$\usym{2715}$ &$\usym{2715}$
				&$\usym{2715}$ 
				&$\checkmark$
				&$\checkmark$
                &$\usym{2715}$
                &$\usym{2715}$
                &$\usym{2715}$
                &$\checkmark$
                &$\usym{2715}$
                \\\cline{1-11}
				
				\textbf{\cite{DBLP:journals/wcl/MohamedHH22}} &$\usym{2715}$ &$\checkmark$ 
				&$\checkmark$ 
                    &$\usym{2715}$
                    &$\usym{2715}$
                    &$\usym{2715}$
				&$\checkmark$
				&$\usym{2715}$
                &$\usym{2715}$
                    &$\usym{2715}$\\
				\cline{1-11}
				\textbf{\cite{song2024enhancing}}  	 
				&$\usym{2715}$
				&$\checkmark$
				&$\usym{2715}$
				&$\checkmark$
				&$\checkmark$
				&$\usym{2715}$
				&$\checkmark$
				&$\usym{2715}$
				&$\usym{2715}$
				&$\usym{2715}$
				\\\cline{1-11}
				\textbf{\cite{DBLP:journals/sj/AdamOWMAML24}}	 
				&$\usym{2715}$
				&$\checkmark$
				&$\checkmark$
				&$\usym{2715}$
				&$\checkmark$
				&$\usym{2715}$
				&$\checkmark$
				&$\usym{2715}$
				&$\usym{2715}$
				&$\usym{2715}$
				\\\cline{1-11}
				\textbf{\cite{DBLP:journals/wcl/WangNTEN23}}
				&$\usym{2715}$
				&$\usym{2715}$
				&$\checkmark$
				&$\usym{2715}$
				&$\checkmark$
				&$\usym{2715}$
				&$\checkmark$
				&$\usym{2715}$
				&$\usym{2715}$
				&$\usym{2715}$
				\\\cline{1-11}
				\textbf{\cite{DBLP:journals/icl/0002TTD0HK23}} 	 
				&$\usym{2715}$
				&$\usym{2715}$
				&$\usym{2715}$
				&$\usym{2715}$
				&$\checkmark$
				&$\usym{2715}$
				&$\checkmark$
				&$\checkmark$
				&$\checkmark$
				&$\usym{2715}$
				\\\cline{1-11}
				
				\textbf{\cite{DBLP:journals/tvt/LiangZDSLW24}}  
				&$\usym{2715}$
				&$\checkmark$
				&$\usym{2715}$
				&$\usym{2715}$
				&$\usym{2715}$
				&$\usym{2715}$
				&$\checkmark$
				&$\usym{2715}$
				&$\usym{2715}$
				&$\usym{2715}$
				\\\cline{1-11}
				\textbf{\cite{DBLP:journals/iotj/TyrovolasMBMTDILK24}}  	 
				&$\usym{2715}$
				&$\checkmark$
				&$\usym{2715}$
				&$\usym{2715}$
				&$\usym{2715}$
				&$\checkmark$
				&$\usym{2715}$
				&$\usym{2715}$
				&$\usym{2715}$
				&$\usym{2715}$
				\\\cline{1-11}
                \textbf{\cite{DBLP:journals/twc/AbouamerM22}}
                &$\usym{2715}$
                &$\checkmark$
                &$\usym{2715}$
                &$\usym{2715}$
                &$\checkmark$
                &$\usym{2715}$
                &$\checkmark$
                &$\usym{2715}$
                &$\usym{2715}$
                &$\usym{2715}$
                \\\cline{1-11}

                \textbf{\cite{DBLP:journals/ojcs/ShehabCKAT22}}
                &$\usym{2715}$
                &$\checkmark$
                &$\usym{2715}$
                &$\usym{2715}$
                &$\usym{2715}$
                &$\usym{2715}$
                &$\checkmark$
                &$\usym{2715}$
                &$\usym{2715}$
                &$\usym{2715}$
                \\\cline{1-11}

                \textbf{\cite{DBLP:journals/wcl/WangLSFS20}}
                &$\usym{2715}$
                &$\checkmark$
                &$\usym{2715}$
                &$\checkmark$
                &$\checkmark$
                &$\usym{2715}$
                &$\usym{2715}$
                &$\usym{2715}$
                &$\checkmark$
                &$\usym{2715}$
                \\\cline{1-11}
                
                
                \textbf{\cite{DBLP:journals/twc/YangCSXSPC22}}  	 
                &$\checkmark$
                &$\checkmark$ 
                &$\usym{2715}$ 
                &$\usym{2715}$
                &$\checkmark$
                &$\usym{2715}$
                &$\checkmark$
                &$\usym{2715}$
                &$\checkmark$
                &$\usym{2715}$
                \\\cline{1-11}
               \textbf{\cite{DBLP:conf/wcnc/WuGLGKK24}}
               &$\checkmark$
               &$\checkmark$
               &$\usym{2715}$
               &$\checkmark$
               &$\checkmark$
               &$\usym{2715}$
               &$\checkmark$
               &$\checkmark$
               &$\checkmark$
               &$\usym{2715}$
               \\\cline{1-11}
               \textbf{\cite{DBLP:conf/vtc/Nasir24}}	 
               &$\usym{2715}$
               &$\checkmark$
               &$\usym{2715}$
               &$\usym{2715}$
               &$\usym{2715}$
               &$\usym{2715}$
               &$\checkmark$
               &$\checkmark$
               &$\checkmark$
               &$\usym{2715}$
               \\\cline{1-11}

                \textbf{\cite{DBLP:journals/wcl/ZhaiDDWY22}}
                &$\usym{2715}$
                &$\checkmark$
                &$\usym{2715}$
                &$\usym{2715}$
                &$\usym{2715}$
                &$\usym{2715}$
                &$\checkmark$
                &$\checkmark$
                &$\checkmark$
                &$\usym{2715}$
                \\\cline{1-11}
                \textbf{\cite{DBLP:conf/icc/MagboolKF23}}
                &$\usym{2715}$
                &$\checkmark$
                &$\usym{2715}$
                &$\usym{2715}$
                &$\checkmark$
                &$\checkmark$
                &$\checkmark$
                &$\usym{2715}$
                &$\checkmark$
                &$\usym{2715}$
                \\\cline{1-11}
                

				\textbf{\cite{DBLP:journals/twc/MaFZGY22}} &$\checkmark$  &$\checkmark$  &$\usym{2715}$ &$\usym{2715}$
                &$\checkmark$
				&$\usym{2715}$
                &$\checkmark$
                &$\usym{2715}$
                &$\usym{2715}$
                &$\usym{2715}$
                \\\cline{1-11}

				\textbf{\cite{DBLP:journals/jsac/LiuLCP21}} &$\usym{2715}$ &$\checkmark$ 
				&$\checkmark$  
				&$\usym{2715}$
				&$\checkmark$
                &$\usym{2715}$
                &$\checkmark$
                &$\usym{2715}$
                &$\usym{2715}$
                &$\usym{2715}$
				\\\cline{1-11}

            \textbf{\cite{DBLP:journals/iotj/NaeemQC23}} &$\checkmark$
				&$\usym{2715}$
				&$\usym{2715}$ 
				&$\usym{2715}$
				&$\checkmark$
                &$\usym{2715}$
                &$\checkmark$
                &$\usym{2715}$
                &$\usym{2715}$
                &$\usym{2715}$
                \\\cline{1-11}
                \textbf{\cite{DBLP:journals/tvt/LinYZXHN24}}  	 
                &$\usym{2715}$
                &$\checkmark$
                &$\usym{2715}$
                &$\usym{2715}$
                &$\usym{2715}$
                &$\usym{2715}$
                &$\checkmark$
                &$\usym{2715}$
                &$\checkmark$
                &$\usym{2715}$
                \\\cline{1-11}
			    \textbf{\cite{DBLP:conf/ictc/ZhaoMLP22}} 	 
				&$\checkmark$
				&$\checkmark$
				&$\usym{2715}$
				&$\usym{2715}$
				&$\usym{2715}$
				&$\usym{2715}$
				&$\usym{2715}$
				&$\usym{2715}$
				&$\usym{2715}$
				&$\usym{2715}$
				\\\cline{1-11}
                
                 This work
				&$\checkmark$
				&$\checkmark$
				&$\checkmark$
				&$\checkmark$
				&$\checkmark$
                &$\checkmark$
                &$\checkmark$
                &$\checkmark$
                &$\checkmark$
                &$\checkmark$
                \\\cline{1-11}
		\end{tabular}}
	\end{center}
 \vspace{-0.6cm}
\end{table*}
\section{Related Work}
\label{sec:Related Work}
\par In this section, we illustrate the differences between the previous works and this work from different perspectives, and the explicit details are highlighted in Table \ref{Reference1}.

\subsection{RIS-enabled Communication Scenarios}

\par There were some previous works considering RIS-assisted wireless communications including multiple RISs or multiple GUs. Specifically, Khisa \emph{et al}. \cite{10109654} considered an RIS-assisted cellular network with a BS, an RIS, and two GUs. With the goal of minimizing the network energy consumption, a joint optimization framework was proposed, subject to the power budget constraints at both of the BS and the relaying node. The authors in \cite{10025789} used cooperative RISs to assist Internet-of-Things (IoT) networks, which could adjust stand-alone or cooperative configurations of the IoT networks.

\par However, in the real scenarios, considering multiple RIS service and multiple GUs is practical, while the abovementioned works only considered either multiple RISs or multiple GUs. Once both of these two factors are considered simultaneously, the complexity of the problem will be increased.

\par \textcolor{black}{In addition, some works considered UAV-RISs-assisted wireless communications. Specifically, Zhao \emph{et al}. in \cite{DBLP:journals/tits/ZhaoSNXGZ24} utilized a UAV-RIS to maximize the available rate between BS and a mobile vehicle. For this purpose, they proposed a position prediction strategy that could adjust UAV-RIS trajectory and phase shifts in time. In \cite{DBLP:conf/globecom/ZhangSB19}, Zhang \emph{et al}. proposed a novel approach that used a UAV-RIS to enhance the  millimeter wave networks. For this purpose, they formulated an optimization problem to maximize the total downlink transmission by adjusting the location and reflection parameter of UAV-RIS. Liu \emph{et al}. in \cite{DBLP:journals/jsac/LiuLC21} considered a UAV-RIS-assisted wireless networks with non-orthogonal multiple access (NOMA) technique to improve the spectrum efficiency, while harvesting energy from millimeter wave signals to power the UAV-RIS. To this end, a decaying deep Q-network was proposed to design the movement, phase shifts of the UAV-RIS system, and power allocation policy of UAV-RIS system. However, only a single UAV-RIS was considered in \cite{DBLP:conf/globecom/ZhangSB19, DBLP:journals/jsac/LiuLC21}. Thus, we utilized multiple UAV-RISs to assist multiple GUs in our work.} 

\subsection{Decision Variables in RIS Systems}

\par Some previous works considered 3D location deployment of RIS, discrete phase shift of RIS or beamforming vector of BS. For example, the authors in \cite{DBLP:journals/wcl/GaoCHY21} optimized the transmit beamforming at BS and discrete phase shifts at RIS to minimize the transmission power of BS in an RIS-assisted wireless communication system. Moreover, the authors in \cite{DBLP:journals/wcl/MohamedHH22} investigated the applications of UAV-RIS for millimeter-wave BS in covering GUs in hotspot areas, and they maximized the available rate of GUs while minimizing the energy consumption of UAV flight from one hotspot to another over the time span of its battery life. \textcolor{black}{Song \emph{et al}. in \cite{song2024enhancing} regarded 2D location deployment of UAV-RIS, the discrete phase shifts of UAV-RIS, and the beamforming vector of BS as their decision variables, so as to achieve emergency coverage requirements and dynamic environments for UAV-RIS. The authors in \cite{DBLP:journals/sj/AdamOWMAML24} considered 3D location deployment of UAV-RIS, the continuous phase shifts of UAV-RIS, and the beamforming vector of BS, and they optimized these decision variables to maximize the secure rate with channel uncertainty constraints. In \cite{DBLP:journals/wcl/WangNTEN23}, the transmit beamforming at the BS, the coefficient matrix and 3D location of the UAV-RIS were optimized to maximize the secure rate. Xiao \emph{et al}. in \cite{DBLP:journals/icl/0002TTD0HK23} investigated a solar-powered UAV-RIS system, where the beamforming vector at BS, phase shifts, 2D hovering location, flying speed, number of reflecting elements, flying time, and hovering time of UAV-RIS are jointed adjusted to maximize the energy efficiency of the system.} 
\par \textcolor{black}{However, in this paper, we considered the 3D location deployment of UAV-RISs, the discrete phase shifts of UAV-RISs, and the beamforming vector of BS as our decision variables, which is more comprehensive. Specifically, the 3D location deployment of UAV-RIS provides a significant advantage over 2D deployment, which allows for a better LoS link and coverage compared to 2D configurations. Additionally, the discrete phase shifts of UAV-RIS are more practical and suitable for the hardware constraints of UAVs, which makes the system more feasible for real-world UAV-RIS deployments, where hardware resources are often limited. In addition, considering beamforming vector enables more precise and flexible signal steering, focusing energy on the GUs while reducing interference in other directions. }

\subsection{Optimization Objectives and Metrics in RIS-assisted Communications}
\par Some previous works only considered a single optimization objective. For instance, \textcolor{black}{the authors in \cite{DBLP:journals/tvt/LiangZDSLW24} aimed to optimize the average available rate in a UAV-RIS-assisted underlay cognitive radio network. The authors in \cite{DBLP:journals/iotj/TyrovolasMBMTDILK24} were dedicated to maximizing the minimum data rate, taking into account mobility, GU scheduling, and UAV-RIS power consumption constraints.} The authors in \cite{DBLP:journals/twc/AbouamerM22} aimed to maximize the total available rate of the uplink and downlink for multiple GUs in an RIS-assisted communication system. The authors in \cite{DBLP:journals/ojcs/ShehabCKAT22} considered an RIS-assisted downlink NOMA scenario to maximize the total available rate of GUs, then proposing a novel deep reinforcement learning (DRL) method. Moreover, Wang \emph{et al}. \cite{DBLP:journals/wcl/WangLSFS20} considered an RIS-assisted NOMA system, and they aimed at minimizing the power consumption by designing the power allocation at the BS and passive beamforming at the RIS jointly. \textcolor{black}{In \cite{DBLP:journals/twc/KhaliliMZJYJ22}, the authors investigated a heterogeneous network supported by dual connectivity with multiple UAV-RISs, wherein minimizing the total transmit power by jointly optimizing the trajectory of UAV-RISs, phase shifts of UAV-RISs, sub-carrier allocation, and active beamformers at each BS. Then, they adopted deep Q-network and successive convex approximation to solve the problem.}
\par \textcolor{black}{However, we considered the minimum available rate over all GUs, the total available rate of all GUs, and the total energy consumption of the system simultaneously. Such a consideration was more comprehensive for actual scenarios, as it addressed the diverse requirements of real-world applications. Specifically, improving the minimum available rate and the total available rate will reduce the complaints from GUs and increase the overall rating of this communication system, respectively, while saving the energy consumption is helpful to control the costs of the system.}

\par Moreover, there were also some works considering several optimization objectives. For example,
\textcolor{black}{a wireless communication network with several cooperative RISs was studied in \cite{DBLP:journals/twc/YangCSXSPC22}, where the authors aimed to maximize the total available rate and communication energy consumption by dynamically scheduling the on-off status of each RIS and optimizing the phase shifts of the RISs. Wu \emph{et al}. in \cite{DBLP:conf/wcnc/WuGLGKK24} considered optimizing the total available rate, UAV flight energy consumption, and communication energy consumption, so as to take over signal strength degradation over long transmission distances and limited spectrum resources. The authors in \cite{DBLP:conf/vtc/Nasir24} investigated a secure and energy-efficient mobile edge computing with UAV-RIS assistance, where they also aimed to optimize the total available rate, UAV flight energy consumption, and communication energy consumption.} Zhai \emph{et al}. \cite{DBLP:journals/wcl/ZhaiDDWY22} also considered a UAV-RIS-assisted mobile edge computing system, where they optimized the energy efficiency of the system by optimizing the communication rate and total energy consumption. Moreover, the authors in \cite{DBLP:conf/icc/MagboolKF23} considered an RIS-assisted millimeter-wave communication system, and they optimized the total available rate, minimum available rate, and communication energy consumption by a two-stage method. 

\par \textcolor{black}{Although \cite{DBLP:journals/twc/YangCSXSPC22, DBLP:conf/wcnc/WuGLGKK24, DBLP:conf/vtc/Nasir24, DBLP:journals/wcl/ZhaiDDWY22, DBLP:conf/icc/MagboolKF23} aimed to optimize multiple objectives in their consideration, we formulated the optimization problem based on Pareto dominance. Such a consideration was practical but challenging. Specifically, these works transformed the multiple optimization objectives into a single optimization objective using a linear weighting method or a quotient method, while introducing Pareto dominance to deal with all optimization objectives is more efficient. The reason is that the algorithm with Pareto dominance can provide several solutions when iteration terminates, and they can be chosen for the decision-makers according to different requirements, which means that the decision-makers do not need to rerun the algorithm when facing different applications. Thus, such a consideration can further enhance the dynamics of the scenario and the scalability of the system, which is more practical.} 

\subsection{Optimization Approaches for RIS-assisted Communications}
\par To solve the complex optimization problem for RIS-assisted wireless communications, many researchers designed several effective algorithms. Specifically, Ma \emph{et al}. \cite{DBLP:journals/twc/MaFZGY22} considered a multi-hop cooperative RISs to maximize the total available rate of all GUs. Since the formulated problem was non-convex and there were interactions among the decision variables, they first decoupled the problem into three subproblems. Then, they adopted the conventional convex optimization methods to solve these subproblems. Liu \emph{et al}. \cite{DBLP:journals/jsac/LiuLCP21} exploited a DRL method to solve RIS deployment, phase shift design, as well as power allocation in an NOMA system. Simulations verified that their algorithm was capable of striking a trade-off between the prediction accuracy and computational complexity. Moreover, the authors in \cite{DBLP:journals/iotj/NaeemQC23} proposed a DRL method empowered by a generative adversarial network. Then, the joint optimization problem for RIS deployment and reflecting beamforming matrix was addressed.
\textcolor{black}{Lin \emph{et al}. in \cite{DBLP:journals/tvt/LinYZXHN24} adopted an improved DRL to maximize the system energy efficiency considering jamming noise and quality of service constraints for maritime users. The authors in \cite{DBLP:conf/ictc/ZhaoMLP22} solved a convex problem to minimize the total number of UAV-RISs while satisfying the user signal-to noise ratio.}

\par \textcolor{black}{However, the conventional convex optimization methods can only solve the convex problems. Furthermore, the DRL methods typically require a substantial number of samples or interactions with the environment in order to learn and acquire optimal policies. This process can be time-consuming and computationally expensive, as it often involves trial and error, and the model may take a long time to converge to a good solution. Additionally, when the application scenario changes such as a shift in network conditions, the decision-makers would need to retrain the model. This retraining process not only increases the execution time but also reduces the adaptability of the algorithm in dynamic environments. Given these challenges, our goal is to design a more efficient algorithm that could handle multiple optimization objectives simultaneously while being robust to changes in the application scenario.}

\section{System Model}
\label{sec:Models}
\par In this section, the system overview and the corresponding models are introduced.

\subsection{System Overview}

\par
\begin{figure*}[htbp]
	\setlength{\abovedisplayskip}{1pt}
	\setlength{\belowdisplayskip}{1pt}
	\setlength{\abovecaptionskip}{1pt}
	\centering{\includegraphics[width=6in]{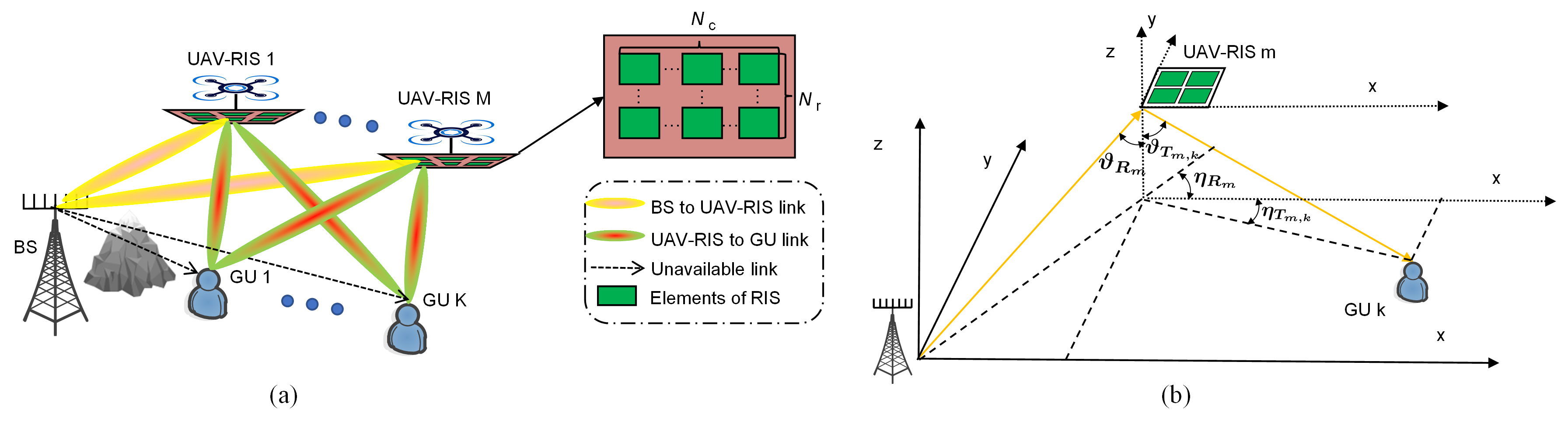}}
	\caption{System structure diagram. (a) A cooperative UAV-RISs-assisted cellular network. (b) Information flow from BS to GU via a UAV-RIS.}
	\label{System_model}
\end{figure*}

 As depicted in Fig. \ref{System_model}, we illustrate the considered cellular network configuration in this work. Specifically, the network consists of a ground BS equipped with $N_{\rm{BS}}$ antennas and $K$ static GUs\footnote{We assume that the locations of GUs are known in advance, since GUs can be equipped with devices capable of determining and transmitting their locations, such as GPS or positioning systems \cite{DBLP:journals/twc/FengXYX22}.}, and each of which is equipped with a single antenna. However, the direct links from the BS to GUs are entirely unavailable. Thus, several cooperative UAV-RISs are employed in a 3D space to reflect the signals from BS to GUs\footnote{We default that the cooperative UAV-RIS system has achieved time synchronization via adopting the existing methods, such as the training and transmission method in \cite{DBLP:journals/jstsp/ZhaoXYWS22}, and hence we do not take extra time synchronization mechanism into account in this work.}. Note that the cooperative UAV-RIS scheme performs better than a single UAV-RIS scheme \cite{DBLP:journals/icl/FaisalADN22} in terms of the energy efficiency of the system, and hence we consider different UAV-RISs to serve several GUs simultaneously. Without loss of generality, we assume that the data size $Q$ for each GU is identical.

\par Specifically, the sets of UAV-RISs and GUs are denoted as $\mathcal{M}=\{1, 2, ..., M\}$ and $\mathcal{K}=\{1, 2, ..., K\}$, respectively. The BS employs a half-wavelength uniform linear array, while each RIS adopts a uniform planar array, comprising $N_{\rm{RIS}}$ passive reflection elements, where $N_{\rm{RIS}}=N_{\rm{r}} \times N_{\rm{c}}$. $N_{\rm{r}}$ and $N_{\rm{c}}$ represent the element numbers along the $y$- and $x$-axes, respectively, as shown in Fig. \ref{System_model}(a).

\par Signals reflected by RISs two or more times can be neglected due to the significant path loss of multi-hop links \cite{DBLP:journals/twc/YangCSXSPC22}. Moreover, we consider a horizontal square area with minimum and maximum ranges of $L_{\min}$ and $L_{\max}$, respectively, and the heights can be adjusted within the range $[Z_{\min}, Z_{\max}]$. It should be noted that all UAV-RISs initially have the position $[0, 0, 0]$ and then fly to their hovering positions at a constant speed for passive reflection. Moreover, we assume that all UAV-RISs are deployed and recalled simultaneously. In other words, the UAV-RISs performing other tasks can be deployed in batches after completing tasks, so as to achieve the opportunistic deployment in 3D space. In the 3D Cartesian coordinate system, the 3D coordinates of the BS, the $k$-th GU, and the $m$-th UAV-RIS are denoted as $\mathbf{q}_{\rm{B}} = [x_{\rm{B}}, y_{\rm{B}}, 0]$, ${\mathbf{q}_{\rm{G}}}_k = [{x_{\rm{G}}}_k, {y_{\rm{G}}}_k, 0]$, and ${\mathbf{q}_{\rm{U}}}_m = [{x_{\rm{U}}}_m, {y_{\rm{U}}}_m, {z_{\rm{U}}}_m]$, respectively. Furthermore, $\mathbf{w}_{\rm{B}}  = [x_{\rm{B}}, y_{\rm{B}}]$, ${\mathbf{w}_{\rm{G}}}_k = [{x_{\rm{G}}}_k, {y_{\rm{G}}}_k]$, and ${\mathbf{w}_{\rm{U}}}_m= [{x_{\rm{U}}}_m, {y_{\rm{U}}}_m]$ represent the horizontal coordinates of the BS, $k$-th GU, and $m$-th UAV-RIS. Then, the transmitted signal at the BS is as follows:
\begin{equation}
\label{transmitted_signal}
\begin{aligned}
	\boldsymbol{s}=\sum_{k=1}^K \boldsymbol{w}_k s_k,
\end{aligned}
\end{equation}
\noindent where $s_k$ is the unit-power information symbol \cite{DBLP:journals/twc/HuangZADY19}, \cite{DBLP:journals/tcom/WeiHAYZD21} and $\boldsymbol{w}_k\in \mathbb{C}^{N_{\rm{BS}}\times 1}$ is the beamforming vector for GU $k\in \mathcal{K}$.

\par The phase shift matrix of $m$-th UAV-RIS can be optimized using a diagonal matrix $\boldsymbol{\Theta}_{m}=\operatorname{diag}\left(\mathrm{e}^{j \theta_{m,1}}, \ldots, \mathrm{e}^{j \theta_{m,n_{m}}}, \ldots, \mathrm{e}^{j \theta_{m, N_{\rm{RIS}}}}\right) \in \mathbb{C}^{N_{\rm{RIS}} \times N_{\rm{RIS}}}$, where $\theta_{m, n_{m}}\in [0, 2\pi)$ and $n_{m}\in \{1, 2, ..., N_{\rm{RIS}}\}$, $\boldsymbol{\Theta}_{m}$ represents the effective phase shifts applied by all passive reflecting elements of $m$-th UAV-RIS. However, the lightweight UAVs limit their load capability, and the size of RISs is restricted, which poses greater challenges for the hardware design \cite{DBLP:conf/iecon/ZhouQ018}. Thus, the discrete phase shift design is more practical than the continuous phase shift design \cite{DBLP:journals/icl/VuK23}. For simplicity, the phase shift of each RIS element is assumed to be one of $C = 2^c$ discrete values, where $c$ represents the number of bits used for quantizing the phase shift levels \cite{DBLP:journals/tcom/AnXGH22}. Specifically, the set of discrete phase shift of each RIS element can be given as follows:
\begin{equation}
\label{C}
\mathcal{C}=\{0, \Delta \theta_{m, n_{m}}, \ldots,(C-1) \Delta \theta_{m, n_{m}}\},
\end{equation}
\noindent where $\Delta \theta_{m, n_{m}} = 2\pi/C$. Then, the received signal at $k$-th GU can be expressed as follows \cite{DBLP:journals/twc/YangCSXSPC22}:
\begin{equation}
\label{recieved_signal}
\begin{aligned}
y_k&=\left(\sum_{m=1}^M \boldsymbol{h}_{k,m}^H \boldsymbol{\Theta}_{m} \boldsymbol{G}_{m}\right) \boldsymbol{s}+n_k,
\end{aligned}
\end{equation}
\noindent where $\boldsymbol{G}_{m}\in \mathbb{C}^{N_{\rm{RIS}}\times N_{\rm{BS}}}$, and $\boldsymbol{h}_{k,m}^H\in \mathbb{C}^{1 \times N_{\rm{RIS}}}$ are the channel matrix from the BS to the $m$-th UAV-RIS and the channel vector from the $m$-th UAV-RIS to $k$-th GU, respectively. Moreover, $n_k$ is the zero mean additive white Gaussian noise (AWGN) with entries of variance $\sigma^2$. 
\par According to Eqs. (\ref{transmitted_signal}) and (\ref{recieved_signal}), the received signal-to-interference-plus-noise ratio (SINR) of $k$-th GU can be expressed as follows \cite{DBLP:journals/jsac/ChenZBZA21}:
\begin{equation}
\label{SINR}
\operatorname{SINR}_k=\frac{\left|\left(\sum\limits_{m=1}^M \boldsymbol{h}_{k,m}^H \boldsymbol{\Theta}_{m} \boldsymbol{G}_{m}\right) \boldsymbol{w}_k\right|^2}{\sum\limits_{i=1, i \neq k}^K\left|\left(\sum\limits_{m=1}^M \boldsymbol{h}_{k,m}^H \boldsymbol{\Theta}_{m} \boldsymbol{G}_{m}\right) \boldsymbol{w}_i\right|^2+\sigma^2}.
\end{equation}
\noindent Then, the available rate of $k$-th GU can be expressed as follows:
\begin{equation}
\label{R_k}
R_{k}=B \log _2\left(1+\operatorname{SINR}_k\right),
\end{equation}
\noindent where $B$ is the bandwidth of the channel.

\subsection{Channel Model}
\par Without loss of generality, we assume that all the channels follow Rician fading, and $\boldsymbol{G}_{m}$ can be expressed as follows \cite{DBLP:journals/twc/MaFZGY22}, \cite{DBLP:journals/wcl/SongZWYT22}:
\begin{equation}
\label{G_m}
\boldsymbol{G}_{m}=\sqrt{\frac{\beta_0}{d_{\rm{B}, \it{m}}}}\left(\sqrt{\frac{\mathcal{A}}{1+\mathcal{A}}} {\boldsymbol{g}_{\rm{LoS}}}_{\rm{B}, \it{m}}+\sqrt{\frac{1}{1+\mathcal{A}}} {\boldsymbol{g}_{\rm{NLoS}}}_{\rm{B}, \it{m}}\right),
\end{equation}
\noindent where $\beta_0$ denotes the reference channel coefficient, $d_{\rm{B}, \it{m}}= \left\|{\mathbf{q}_{\rm{U}}}_m-\mathbf{q}_{\rm{B}}\right\|$ is the distance between BS and $m$-th UAV-RIS, $\mathcal{A}$ denotes Rician factor, and ${\boldsymbol{g}_{\rm{LoS}}}_{\rm{B}, \it{m}}$ and ${\boldsymbol{g}_{\rm{NLoS}}}_{\rm{B}, \it{m}}$ are the LoS and non-LoS (NLoS) components of the channel. When the height of UAV-RIS is high enough, the channel can be regarded as LoS channel \cite{DBLP:journals/tcom/Al-JarrahAAIA21}. As the Rician factor increases, the channel coefficients become more dependent on the free-space path loss. When the value of the Rician factor satisfies $\mathcal{A}\geq 20$, $\sqrt{{\mathcal{A}}/{(1+\mathcal{A})}} {\boldsymbol{g}_{\rm{LoS}}}_{\rm{B}, \it{m}} \gg \sqrt{{1}/{(1+\mathcal{A})}} {\boldsymbol{g}_{\rm{NLoS}}}_{\rm{B}, \it{m}}$, and then the channel can be approximated as LoS channel. Thus, Eq. (\ref{G_m}) can be simplified as follows:
\begin{equation}
\label{simplified_G_m}
\boldsymbol{G}_{m}=\sqrt{\frac{\beta_0}{d_{\rm{B}, \it{m}}^2}}\left(\sqrt{\frac{\mathcal{A}}{1+\mathcal{A}}} {\boldsymbol{g}_{\rm{LoS}}}_{\rm{B}, \it{m}}\right),
\end{equation}

\noindent where ${\boldsymbol{g}_{\rm{LoS}}}_{\rm{B}, \it{m}}$ can be further expressed as follows \cite{DBLP:journals/twc/MaFZGY22}:
\begin{equation}
\label{g_{B, m, k L o S}}
{\boldsymbol{g}_{\rm{LoS}}}_{\rm{B}, \it{m}}=\mathbf{a}_{R_{m}}\left(\eta_{R_{m}}, \vartheta_{R_{m}}\right) \mathbf{a}_{T}^H\left(\vartheta_{R_{m}}\right),
\end{equation}
\noindent where $\mathbf{a}_{R_{m}}\left(\eta_{R_{m}}, \vartheta_{R_{m}}\right) \in \mathbb{C}^{N_{\rm{RIS}}\times 1}$ is the receiving array response of $m$-th UAV-RIS, and $\mathbf{a}_{T}^H\left(\vartheta_{R_{m}}\right)\in \mathbb{C}^{N_{\rm{BS}}\times 1}$ is the transmit array response of the BS. Specifically, $\mathbf{a}_{R_{m}}$ can be expressed as follows \cite{DBLP:journals/twc/MaFZGY22}:
\begin{equation}
\label{{a}_{m, k}}
\begin{aligned}
\mathbf{a}_{R_{m}}= & {\left[1, \ldots, \mathrm{e}^{\frac{j 2 \pi\left(N_{\rm{c}}-1\right) d \phi_{R_{m}}}{\lambda}}\right]^T} \\
& \otimes\left[1, \ldots, \mathrm{e}^{\frac{j 2 \pi\left(N_{\rm{r}}-1\right) d \Omega_{R_{m}}}{\lambda}}\right]^T,
\end{aligned}
\end{equation}
\noindent where $d=\lambda/2$ is the distance between two adjacent elements on one RIS, and $\lambda$ is the signal wavelength. $\phi_{R_{m}} = \sin(\vartheta_{R_{m}})\cos(\eta_{R_{m}})$ and $\Omega_{R_{m}} = \sin(\vartheta_{R_{m}})\sin(\eta_{R_{m}})$ are the angle parameters. \textcolor{black}{It can be seen from Fig. \ref{System_model}(b) that $\vartheta_{R_{m}} = \arcsin{\frac{\left\|{\mathbf{w}_{\rm{U}}}_m-\mathbf{w}_{\rm{B}}\right\|}{d_{\rm{B}, \it{m}}}}$ is the zenith angle of arrival (AoA), and $\eta_{R_{m}}=\arccos{\frac{\left\|{x_{\rm{U}}}_m-{x_{\rm{B}}}\right\|}{\left\|{\mathbf{w}_{\rm{U}}}_m-\mathbf{w}_{\rm{B}}\right\|}}$ is the azimuth AoA at $m$-th UAV-RIS. Similarly, we can have the transmit array response $\mathbf{a}_{T}\left(\vartheta_{R_{m}}\right) = \left[1, \mathrm{e}^{j \pi \sin \vartheta_{R_{m}}}, \ldots, \mathrm{e}^{j \pi(N_{\rm{BS}}-1) \sin \vartheta_{R_{m}}}\right]^T$,} in which $\vartheta_{R_{m}}$ is the angle of departure (AoD) from BS to $m$-th UAV-RIS.

\par Then, we can also obtain $\boldsymbol{h}_{k,m}$ following the similar process, which can be expressed as follows:
\begin{equation}
\label{{h}_{k,m}}
\boldsymbol{h}_{k,m}=\sqrt{\frac{\beta_0}{d_{m, k}^2}}\left(\sqrt{\frac{\mathcal{A}}{1+\mathcal{A}}} {\boldsymbol{g}_{\rm{LoS}}}_{\rm{R}, \it{m}, \it{k}}\right),
\end{equation}
\noindent \textcolor{black}{where $d_{\rm{B}, \it{m}}=\left\|{\mathbf{q}_{\rm{U}}}_m-{\mathbf{q}_{\rm{G}}}_k\right\|$ is the distance between $m$-th UAV-RIS and $k$-th GU, ${\boldsymbol{g}_{\rm{LoS}}}_{\rm{R}, \it{m}, \it{k}} = \mathbf{a}_{T_{m, k}}^T\left(\eta_{T_{m, k}}, \vartheta_{T_{m, k}}\right)$, and $ \mathbf{a}_{T_{m, k}}$ represents the transmit array response of $m$-th UAV-RIS. Following the similar definition $\phi_{T_{m, k}} = \sin(\vartheta_{T_{m,k}})\cos(\eta_{T_{m, k}})$ and $\Omega_{T_{m, k}} = \sin(\vartheta_{T_{m, k}})\sin(\eta_{T_{m, k}})$ with AoDs $\vartheta_{T_{m,k}} = \arcsin{\frac{\left\|{\mathbf{w}_{\rm{U}}}_m-{\mathbf{w}_{\rm{G}}}_k\right\|}{d_{m, k}}}$ and $\eta_{T_{m, k}} = \arcsin{\frac{\left\|{y_{\rm{U}}}_m-{y_{\rm{G}}}_k\right\|}{\left\|{\mathbf{w}_{\rm{U}}}_m-{\mathbf{w}_{\rm{G}}}_k\right\|}}$,} we have $\mathbf{a}_{T_{m, k}}$ as follows: 
\color{black}
\begin{equation}
\label{{a}_{T_{m, k}}}
\begin{aligned}
\mathbf{a}_{T_{m, k}}\left(\eta_{T_{m, k}}, \vartheta_{T_{m, k}}\right)= & {\left[1, \ldots, \mathrm{e}^{\frac{j 2 \pi\left(N_{\rm{c}}-1\right) d \phi_{T_{m, k}}}{\lambda}}\right]^T}\\
& \otimes\left[1, \ldots, \mathrm{e}^{\frac{j 2 \pi\left(N_{\rm{r}}-1\right) d \Omega_{T_{m, k}}}{\lambda}}\right]^T.
\end{aligned}
\end{equation}
\subsection{Energy Consumption Model}
\par As mentioned previously, all UAV-RISs are first deployed from their initial position to hovering positions before reflecting signals. Therefore, the total energy consumption can be divided into two parts: the energy consumption during location deployment and the energy consumption during hovering. The latter includes the energy consumed by the UAV-RISs against gravity, the energy consumption by the BS for transmission, the circuit energy consumption by both the BS and GUs, and the energy consumption for reflection by the RIS. Assuming that UAV-RISs fly at a constant velocity $V$, then the energy consumption during deployment can be expressed as follows \cite{DBLP:journals/pieee/ZengWZ19}:
\begin{equation}
\label{Deploying_energy}
\begin{split}
E_{\rm{pro}}&\approx\sum_{m=1}^{M}\left({P_{\rm{pro}}}_{m} {T_{\rm{pro}}}_{m} +M_{\rm{UR}}g({z_{\rm{U}}}_m-{z_{\rm{U}}}_0)\right)\\&+M\frac{M_{\rm{UR}}({V}^2-{V_0}^2)}{2}
\end{split}
\end{equation}
\noindent \textcolor{black}{where ${T_{\rm{pro}}}_{m} = d_{\rm{B}, \it{m}}/V$ is the flight time of the $m$-th UAV-RIS, $M_{\rm{UR}}$ is the mass and $g$ is the gravitational factor, ${{z}_{\rm{U}}}_0$ is the initial height of $m$-th UAV-RIS}, and $V_0$ is an initial velocity. Moreover, ${P_{\rm{pro}}}_{m}$ is the propulsion power of $m$-th UAV-RIS, which is given as follows \cite{DBLP:journals/twc/ZengXZ19}:
\begin{equation}
\label{UAV-2D-Power}
\begin{split}
{P_{\rm{pro}}}_{m}(V)&=P_{\rm{B}}\left(1+\frac{3{V}^2}{U_{\rm{tip}}^{2}}\right)+P_{\rm{I}}\left(\sqrt{1+\frac{{V}^4}{4v_{0}^{4}}}-\frac{{V}^2}{2v_{0}^{4}}\right)^{\frac{1}{2}}\\
	&+\frac{1}{2}d_{0}\rho sAV^{3},
\end{split}
\end{equation}
\noindent where $P_{\rm{B}}$, $P_{\rm{I}}$, $U_{\rm{tip}}$, $v_{0}$, $d_{0}$, $\rho$, $s$ and $A$ are constant parameters related to UAV, which can be found in \cite{DBLP:journals/twc/ZengXZ19}. Note that when $V=0$, $P_{\rm{B}}+P_{\rm{I}}$ is the power required to overcome gravity.

\par To analyze the energy consumption during hovering, we first calculate the transmission time $T_k$ of $k$-th GU, which is given by $T_k = Q/R_k$. Since all UAV-RISs are assumed to be deployed and recalled simultaneously, the hovering time of the UAV-RISs is determined by the maximum transmission time of $k$-th GU, denoted as $T_{\rm{hov}}=\max\{T_k\}$. The energy consumption during hovering can then be expressed as follows:
\begin{equation}
\label{Hovering_energy}
\begin{split}
E_{\rm{hov}} = \left(\sum_{m=1}^{M}(P_{\rm{B}}+P_{\rm{I}}) + P_{\rm{com}}\right)T_{\rm{hov}},
\end{split}
\end{equation}
\noindent \textcolor{black}{where $P_{\rm{com}}$ is the communication power during hovering,} which can be expressed as follows \cite{DBLP:journals/twc/YangCSXSPC22}: 
\color{black}
\begin{equation}
\label{P_{com}}
\begin{aligned}
P_{\rm{com}}& =\underbrace{\sum_{k=1}^K \frac{\boldsymbol{w}_k^H \boldsymbol{w}_k}{\mu}}_{\text {transmit power of BS }}+\underbrace{P_{\mathrm{BS}}}_{\text {circuit power of BS }} \\
& +\underbrace{\sum_{k=1}^K P_k}_{\text {circuit power of all GUs }}+\underbrace{\sum_{m=1}^M N_{\rm{RIS}} P_{\mathrm{R}}}_{\text{power consumption of all RIS elements}},
\end{aligned}
\end{equation}
\noindent where $\mu$ is the power amplifier efficiency of BS, $P_{\mathrm{BS}}$ is the circuit power consumption of BS, $P_k$ is the circuit power consumption of $k$-th GU, and $P_{\mathrm{R}}$ is the power consumption of each reflecting element in the RIS.
\section{Formulated EEComm-MOF}
\label{sec:Problem Formulation and Analysis}

\par In the real world, the cases usually contain three optimization objectives, which are maximizing the minimum available rate, maximizing the total available rate, and minimizing the total energy consumption of the system. \textcolor{black}{First, maximizing the minimum available rate can ensure that the worst-performing GU is guaranteed for a satisfactory service level, which is particularly relevant in mission-critical applications, such as disaster rescue, where a minimum level of performance must be guaranteed to ensure fairness among GUs. Second, maximizing the total available rate focuses on improving overall throughput in the system, which makes GUs receive as much information (e.g. video) as possible. Third, minimizing the total energy consumption can influence the lifetime of the service, preventing the system from terminating service due to energy exhaustion.} Thus, these three optimization objectives should be jointly optimized, and the details are as follows.

\noindent \emph{\textbf{Optimization objective 1: maximize the minimum available rate over all GUs.}} To ensure the fair service of the UAV-RIS system, we aim to optimize the minimum available rate over all GUs. Hence, the first objective function can be formulated as follows: 
\begin{equation}
\label{f_1}
\begin{split}
	f_1 = \min \{R_k\}.
\end{split}
\end{equation}
\noindent \emph{\textbf{Optimization objective 2: maximize the total available rate of all GUs.}} The total available rate of all GUs reflects the system capacity. Thus, the second objective function can be expressed as follows: 
\begin{equation}
\label{f_2}
\begin{split}
	f_2 = \sum_{k=1}^K R_k.
\end{split}
\end{equation}
\noindent \textbf{\emph{Remark 1.}} \textcolor{black}{Considering the abovementioned two optimization objectives in a multi-objective optimization framework is meaningful, and the reasons are as follows. Since the these two optimization objectives are conflict (the detailed analysis is shown in Appendix C), we can obtain the Pareto front (PF) for these two optimization objectives. The PF offers a range of trade-offs between these two optimization objectives, which is essential in providing a flexible and adaptable solution. Thus, such a formulation can enhance the portability for the optimization framework to different deployment scenarios. Specifically, when extending EEComm-MOF to mountain disaster rescue, ensuring minimum available rates for each GU is crucial to maintain communication, even at the cost of some system throughput, especially for GUs at the network edge that need critical information. In this case, maximizing the minimum available rate becomes essential to ensure that each GU is served during an emergency. Instead, when EEComm-MOF is extended to urban communication scenario with a high concentration of GUs, total network performance and throughput are often prioritized, as GUs in urban environments typically have more robust signal coverage, although the communication fairness is still important to avoid significant disparities in service quality. By formulating these two optimization objectives in a multi-objective optimization framework, decision-makers can obtain a PF that contains a range of feasible solutions. These solutions provide decision-makers with the flexibility to select the optimal trade-off based on the specific scenario. In other words, decision-makers can dynamically choose the solution that best matches the unique requirements of each scenario without rerunning the algorithm, providing a more adaptable solution for the scenario.} \\
\noindent\emph{\textbf{Optimization objective 3: minimize the total energy consumption of the UAV-RIS system.}} The total energy consumption of the UAV-RIS system contains the energy consumed for the deployment and hovering. Thus, the third objective function can be formulated as follows: 
\begin{equation}
\label{f_3}
\begin{split}
	f_3 = E_{\rm{pro}} + E_{\rm{hov}}.
\end{split}
\end{equation}
\par Since the first and second optimization objectives are to obtain the maximums, while the third optimization objective is to obtain the minimum, we take the negative values of the first and second optimization objectives to unify the optimization direction. Thus, the ultimate EEComm-MOF can be formulated as follows:
\begin{subequations}
	\label{En_eff_com}
	\begin{align}
	{\underset{\mathbf{\boldsymbol{\theta}, \boldsymbol{w}, \mathbf{q}_{U}}}{\text{min}}}  \quad   F&=	\left <-f_{1}, -f_{2}, f_{3} 	\right >\\
	\text{s.t.} \qquad & L_{\min} \leqslant {x_{\rm{U}}}_m \leqslant  L_{\max}, \forall m \in \mathcal{M},\\
	& L_{\min} \leqslant  {y_{\rm{U}}}_m \leqslant  L_{\max}, \forall m \in \mathcal{M},\\
	& Z_{\min} \leqslant  {z_{\rm{U}}}_m \leqslant  Z_{\max}, \forall m \in \mathcal{M},\\
	& \theta_{m, n_{m}}\in \mathcal{C}, \forall n_{m}\in \{1, 2, ..., N_{\rm{RIS}}\}, \forall m\in \mathcal{M},\\    
	&\boldsymbol{w}^H \boldsymbol{w} \leq P_{\max},
	\end{align}
\end{subequations}
\noindent \textcolor{black}{where $\left <\cdot \right >$ is to consider all three optimization objectives using Pareto dominance, instead of using linear weighting or penalty functions to transform all objectives into a single optimization objective.} $\boldsymbol{\theta}=\left[\theta_{1, 1}; \ldots; \theta_{1, N_{\rm{RIS}}}; \ldots; \theta_{M, N_{\rm{RIS}}}\right]$, and the dimension of $\boldsymbol{\theta}$ is $M \times N_{\rm{RIS}}$. $\boldsymbol{w}=\left[\boldsymbol{w}_1; \ldots; \boldsymbol{w}_K\right]$. $\mathbf{q}_{\rm{U}}= \left[{\mathbf{q}_{\rm{U}}}_1; \ldots;{\mathbf{q}_{\rm{U}}}_M\right]$, and $P_{\max}$ is the maximum transmit power of BS. Eqs. (\ref{En_eff_com}b)-(\ref{En_eff_com}d) limit the flight range of UAV-RISs. The phase shift constraint for each reflecting element is provided in Eq. (\ref{En_eff_com}e), and Eq. (\ref{En_eff_com}f) represents the transmit power constraint. \textcolor{black}{As can be seen, the decision variables are coupled, which means that the optimization of one decision variable affects the optimization of others. Specifically, the 3D location deployment of the UAV-RIS system influences the channel conditions between the UAV-RIS and BS, which in turn determines the optimal beamforming vector of the BS. Additionally, the beamforming vector affects the signal quality at the GUs, which requires adjusting the UAV-RIS phase shifts to optimize the transmission. Thus, changing one variable impacts the other variables, creating a coupling effect. This mutual interaction increases the complexity of EEComm-MOF, since solving for one variable without considering the others is not possible, thus requiring joint optimization of all the decision variables in an integrated manner.} 

\begin{proposition}
	\label{proposition:NP-hard}
	The formulated EEComm-MOF is NP-hard.
\end{proposition}
\emph{Proof}: Please see Appendix A. $\hfill\blacksquare$

\begin{proposition}
	\label{proposition:non-convex}
	The formulated EEComm-MOF is non-convex.
\end{proposition}
\emph{Proof:}  Please see Appendix B.  $\hfill\blacksquare$\\
 
\noindent \textbf{\emph{Remark 2.}} There are trade-offs among the three optimization objectives of EEComm-MOF, and the detailed analysis is shown in Appendix C.

\noindent \textbf{\emph{Remark 3.}} EEComm-MOF is a large-scale optimization problem, and the details are analyzed in Appendix D.

\section{Algorithm for EEComm-MOF}
\label{sec:Algorithm}
\par \textcolor{black}{Due to the complexity of the formulated EEComm-MOF, it is challenging to obtain the optimal solution in polynomial time. Solving this kind of problems can be roughly divided into three categories, which are the conventional convex optimization methods, DRL, and evolutionary multi-objective optimization algorithms. First, owing to the non-convexity of the EEComm-MOF, conventional convex optimization methods are not applicable. The key characteristic of non-convex optimization problems is that the objective function or constraints may have multiple local optimums, which makes it impossible to guarantee that conventional convex optimization methods will find the global optimum. In this context, conventional convex optimization methods, such as gradient descent or Lagrangian dual methods, typically rely on convexity assumptions to ensure convergence and optimality. However, for EEComm-MOF, these methods may fail to provide effective solutions and can get stuck in local optimum. Second, there are two reasons that we do not choose DRL to solve EEComm-MOF: One reason is that DRL will convert the optimization objectives as a single reward function \cite{DBLP:journals/tcom/PanLSFLY23}, while it is difficult to determine the weight ratio due to the trade-offs among optimization objectives. Such a conversion will impact the optimization direction of the problem. Another reason is that DRL is typically employed in scenarios with continuous time slots \cite{DBLP:journals/iotj/FengWHY24}, allowing UAV-RISs to make decisions through real-time training. However, the formulated EEComm-MOF is a problem with a moment, where the obtained solution can be used for a while. In this case, using DRL would introduce unnecessary overhead due to the training process, which is not suitable for the considered scenario.}
\par \textcolor{black}{Evolutionary multi-objective optimization algorithms, such as non-dominated sorting genetic algorithm-II (NSGA-II) \cite{DBLP:journals/tec/DebAPM02}, are a kind of algorithms based on iterations, and they are suitable for solving EEComm-MOF. First, evolutionary multi-objective optimization algorithms can solve the non-convex and constraint optimization problem in polynomial time \cite{zhang2024uav}. Second, evolutionary multi-objective optimization algorithms use PF to judge the quality of a solution in a multi-objective optimization problem, which means that the decision-makers can ultimately obtain a Pareto set and then select one proper solution according to the scenario. Once the requirements change, the decision-makers only need to re-select the solution instead of rerunning the algorithm, which enhances the transferability of the algorithm. Finally, the formulated EEComm-MOF is a deployment optimization problem based on a single time-slot, and the evolutionary multi-objective optimization algorithms are classic random search method with strong robustness and global search capabilities, which can solve the optimization problem at the single time-slot.}
\par Among evolutionary multi-objective optimization algorithms, NSGA-II is chosen as the basic framework to solve EEComm-MOF, and the motivations are as follows.
\color{black}
\begin{itemize}
	\item Capability to Solve Large-scale Optimization Problem: NSGA-II has a higher computational efficiency when dealing with large-scale optimization problems. According to Remark 3, EEComm-MOF is a large-scale optimization problem, and the computational efficiency of NSGA-II is relatively high \cite{DBLP:journals/tec/ZhangTCJ15}, which enables us to solve more intricate multi-objective optimization problems in a reasonable time, ensuring the practicality of the algorithm.
	\item  Crowding Distance Sorting for Maintaining Population Diversity: NSGA-II uses crowding distance to maintain the diversity of the solution set and avoid solutions from becoming overly concentrated. Due to the mutual coupling among decision variables, EEComm-MOF exists a large solution space. The crowding distance sorting enables NSGA-II to select solutions that are evenly distributed along the PF \cite{DBLP:journals/tec/YueQL18}, ensuring the transferability of the algorithm and providing more practical solutions for real-world deployment.
	\item Simple and efficient implementation: NSGA-II is easy to implement with only a few parameters to tune, such as population size and crossover/mutation rates. This simplicity is especially beneficial for practical applications like UAV-RISs-assisted cellular network, where computational resources are limited. 
\end{itemize}
\color{black}
\par In the following, the conventional NSGA-II is briefly introduced.

\subsection{Conventional NSGA-II}

\par The conventional NSGA-II is a typical evolutionary multi-objective optimization algorithm, whose basic concept is similar to genetic algorithm (GA) \cite{DBLP:journals/tcom/ZhiPRW22}. Specifically, NSGA-II employs the chromosome to represent a solution, and then utilizes the crossover and mutation to generate offspring solutions. Different from GA, NSGA-II is specifically designed for multi-objective optimization problems, and hence it is difficult to judge the quality of solutions by directly comparing the value of single objective function. Instead, Pareto optimality concept is introduced.

\par
\begin{figure*}[htbp]
		\setlength{\abovedisplayskip}{1pt}
	\setlength{\belowdisplayskip}{1pt}
	\setlength{\abovecaptionskip}{1pt}
	\centering{\includegraphics[width=6.5in]{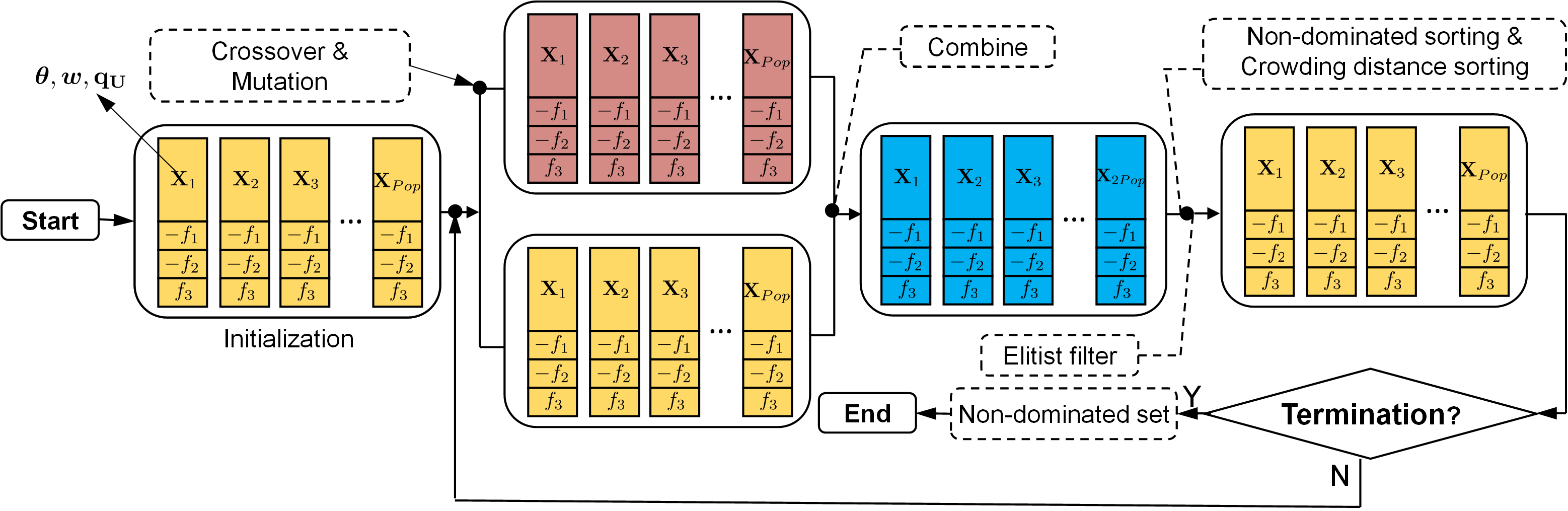}}
	\caption{Outline of NSGA-II.}
	\label{Outline_NSGA_II}
\end{figure*}

 Moreover, NSGA-II incorporates an elitist mechanism including non-dominated sorting and a crowding distance sorting to maintain the population size, which can be described in detail as follows: first, parent population $\mathcal{P}_{it}$ and offspring population $\mathcal{S}_{it}$ will be combined to form a new population $\mathcal{N}_{it}$, and then a non-dominated sorting mechanism ranks $\mathcal{N}_{it}$ into different fronts $\{\mathcal{F}_{1}, \mathcal{F}_{2}, ...\}$ according to their non-domination levels. Second, the solution with the top non-dominated levels will be transplanted into the next generation population until the next generation population size is not less than the preset value $Pop$ for the first time. If the next generation population size is larger than $Pop$ for the first time, NSGA-II will determine which solutions to exclude based on the crowding distance. The explicit outline of NSGA-II is shown in Fig. \ref{Outline_NSGA_II}.
\subsection{Proposed INSGA-II-CDC}
\par The conventional NSGA-II cannot handle the discrete and complex solutions due to its unique crossover and mutation operators. \textcolor{black}{Accordingly, we propose INSGA-II-CDC for a cooperative UAV-RISs-assisted cellular network, which means that it is more suitable for solving the formulated EEComm-
MOF. Specifically, we embed an opposition-based learning operator into continuous solution processing mechanism of NSGA-II to exploit the search space for better deployed locations of UAV-RISs. Then, a discrete solution processing mechanism and a complex solution processing mechanism are embedded into NSGA-II to deal with discrete and complex solutions, respectively, so as to learn the discrete phase shifts of UAV-RISs from population, and efficiently update the beamforming vector of BS.} Assume $G_{\max}$ is the maximum iteration. Then, the overall structure of INSGA-II-CDC is illustrated in Algorithm \ref{INSGA-II-CDC}, and the details of the continuous solution processing mechanism, discrete solution processing mechanism, and complex solution processing mechanism are as follows.

\begin{algorithm}[htb]
	\caption{INSGA-II-CDC}
	\label{INSGA-II-CDC}
	{\textbf{Input:} $Pop$, $G_{\max}$, etc.\\
		\textbf{Output:} The non-dominated set $\mathcal{F}_{1}$.\\	
		$\mathcal{P}_{0}\Leftarrow \varnothing$;\\
		Initialize the UAV-RIS locations in the boundary randomly;\\
		Initialize the phase shifts of UAV-RISs using Algorithm \ref{LargePhase};\\
		Initialize the beamforming vector randomly and then normalize it using Algorithm \ref{Normalization};\\
		\For{$it$= $1$ to $G_{\max}$}{
				Update UAV-RIS locations to generate the offspring ${\mathcal{S}_1}_{it}$ using crossover \cite{DBLP:journals/tec/DebAPM02} and mutation \cite{DBLP:journals/tec/DebAPM02};\\
				Modify ${\mathcal{S}_1}_{it}$ using Eq. (\ref{OBL});\\
				Update phase shifts of UAV-RISs to generate the offspring ${\mathcal{S}_2}_{it}$ using Algorithms \ref{LargePhase} and \ref{phase_learning};\\
				Update beamforming vector and then normalize it to generate the offspring ${\mathcal{S}_3}_{it}$ according to Eq. (\ref{comupdate}) and Algorithm \ref{Normalization};\\
				$\mathcal{P}_{it+1}\Leftarrow \mathcal{P}_{it} \cup{\mathcal{S}_1}_{it}\cup{\mathcal{S}_2}_{it}\cup{\mathcal{S}_3}_{it}$;\\
				Calculate the objective functions according to Eqs. (\ref{transmitted_signal})-(\ref{En_eff_com});\\
				Execute elitist filter to maintain the population size;\\			
		}
	}	
\end{algorithm}

\subsubsection{Continuous Solution Processing Mechanism}

\par The continuous solution of the formulated problem is the deployed locations of UAV-RISs. Note that the location deployment has to satisfy the boundary constraints, i.e., Eqs. (\ref{En_eff_com}b)-(\ref{En_eff_com}d). In the conventional NSGA-II, the algorithm utilizes crossover and mutation operators to generate new solutions. However, when using the mutation operator, there is a risk of the UAV-RISs flying outside the boundary. Facing this situation, the conventional NSGA-II only simply set the locations at the boundaries. Obviously, such a setting is unreasonable because UAV-RISs may be far from some GUs, adversely affecting the first optimization objective $f_1$. Thus, we introduce an opposition-based learning operator when UAV-RISs are out of the boundaries \cite{DBLP:journals/tnse/LongZDPOX23}, which can be expressed as follows:
\begin{subequations}
	\label{OBL}
\begin{align}
		{x_{\rm{U}}}_m &=  \left \{ 
            \begin{array}{ll}
            2L_{\max}-{x_{\rm{U}}}_m, \quad \text{if}\quad {x_{\rm{U}}}_m\geq L_{\max}. \\
		  L_{\min}-{x_{\rm{U}}}_m, \quad \text{if}\quad {x_{\rm{U}}}_m\leq L_{\min}. \\
            \end{array}\right.\\     
		{y_{\rm{U}}}_m &=  \left \{ 
            \begin{array}{ll}
            2L_{\max}-{y_{\rm{U}}}_m, \quad \text{if}\quad {y_{\rm{U}}}_m\geq L_{\max}. \\
            L_{\min}-{y_{\rm{U}}}_m, \quad \text{if}\quad {y_{\rm{U}}}_m\leq L_{\min}. \\
            \end{array}\right.\\
		{z_{\rm{U}}}_m &= \left \{ 
            \begin{array}{ll}
            2Z_{\max}-{z_{\rm{U}}}_m, \quad \text{if}\quad {z_{\rm{U}}}_m\geq Z_{\max}. \\
		Z_{\min}-{z_{\rm{U}}}_m, \quad \text{if}\quad {z_{\rm{U}}}_m\leq Z_{\min}.
            \end{array}\right.
\end{align}
\end{subequations}
\noindent It can be seen from Eq. (\ref{OBL}), more potential deployed locations of UAV-RISs can be exploited instead of only the boundaries, when the UAV-RISs fly outside the boundary. Thus, the search efficiency of the algorithm is enhanced.

\subsubsection{Discrete Solution Processing Mechanism}
\par According to the constraint Eq. (\ref{En_eff_com}e), the set of permissible phase shifts can be viewed as a set of points, where the cardinality of the set is determined by the number of bits used for quantizing the phase shift levels $c$. Moreover, the dimension of discrete solution is $M\times N_{\rm{RIS}}$. With increases in the number of UAV-RISs and the number of RIS elements, the dimension will be increased. Thus, we propose two phase shift adjustment operators to adaptively and iteratively update the phase shifts. Assume that $\text{randi}[C]$ is to randomly generate an integer from $\{0, 1, \ldots, C-1 \}$. Then, Algorithm \ref{LargePhase} presents the first operator, known as the random phase shift operator. Due to the nature of randomness, random search is more likely to explore different regions during the search and thus more likely to find a globally optimal solution.
\begin{algorithm}[htb]
	\caption{Random phase shift operator}
	\label{LargePhase}
	\textbf{Input:} $C=2^c$.\\
	\For{$pop$= $1$ to $Pop$}{
		\For{$m$= $1$ to $M$}{
			\For{$n_m$= $1$ to $N_{\rm{RIS}}$}{
				$\theta_{m, n_m} = \frac{2\pi}{C} \text{randi}[C]$;
			}
		}		
	}
	Record the generated offspring as ${{\mathcal{S}_2}_{it}}^{\prime}$.
\end{algorithm}
\par However, such randomness will have no learning ability, and will not adjust future search directions based on the results of previous searches, which can lead to waste of search time. Thus, we propose a phase shift learning operator to update a discrete solution, whose main idea is to learn from the best solution currently obtained. Specifically, assume that $n_{\mathcal{F}_1}$ is the number of solutions in $\mathcal{F}_1$ with the maximum crowding distance. We first select a discrete solution from $\mathcal{F}_1$ according to crowding distance, and then use this discrete solution to construct a new offspring ${{\mathcal{S}_2}_{it}}^{\prime\prime}$. The detailed process is given  in Algorithm \ref{phase_learning}.

\begin{algorithm}[htb]
	\caption{Phase learning operator}
	\label{phase_learning}
	\textbf{Input:} $\mathcal{F}_1$ and ${{\mathcal{S}_2}_{it}}^{\prime}$.\\
	\eIf {$n_{\mathcal{F}_1} =1 $}{
		Select the only discrete solution with the maximum crowding distance;\\
		
	}(\tcc*[f]{$n_{\mathcal{F}_1} >1 $})
	{Select a discrete solution randomly from $\mathcal{F}_1$ with maximum crowding distance;\\}
	\For{$pop$= $1$ to $Pop$}{
		Replace the discrete solution of $pop$-th solution with the selected discrete solution;\\
	}
	Record the generated offspring as ${{\mathcal{S}_2}_{it}}^{\prime\prime}$;\\
	${\mathcal{S}_2}_{it}= {{\mathcal{S}_2}_{it}}^{\prime}\cup {{\mathcal{S}_2}_{it}}^{\prime\prime}$.
\end{algorithm}

\subsubsection{Complex Solution Processing Mechanism}
\par As shown in Eq. (\ref{En_eff_com}), the beamforming vector is a complex solution, and it must satisfy the constraint Eq. (\ref{En_eff_com}f). However, the conventional NSGA-II can only deal with the continuous solution. Although we can convert each beamforming vector to transmit power by calculating the Euclidean norm, so that the complex solution is converted to the continuous form which can be coped with by the algorithm, the important phase information of the beamforming vector will be missing. Thus, utilizing crossover and mutation operators to update such a continuous form will decrease the search efficiency. Not only that, using mutation will also lead to violation of Eq. (\ref{En_eff_com}f), and the conventional method of dealing with out-of-bounds is difficult to effectively deal with both the amplitude information and phase information. In addition, using a penalty function to deal with Eq. (\ref{En_eff_com}f) may change the search direction, which can further reduce the search efficiency. Thus, it is also necessary to introduce an efficient complex solution processing mechanism for solving it. 

\par Multi-objective particle swarm optimization (MOPSO) is an efficient approach and can trace individual historical optimal solution $Pbest$ \cite{DBLP:journals/tec/CoelloPL04}. Thus, the solution update method in MOPSO is embedded to update the beamforming vector. Specifically, an external archive is used for storing and retrieving the non-dominated solutions obtained. Then, a roulette-wheel method is used to select a solution $Gbest$ from the external archive \cite{DBLP:journals/eswa/MirjaliliSMC16}. To save space complexity, we simply use $\mathcal{F}_1$ instead of an external archive in this work. Afterward, update the beamforming vector as follows: 
\begin{subequations}
	\label{comupdate}
	\begin{align}
		Ve_{i,k}=&\epsilon*Ve_{i,k}+c_1*r_1*({Pbest}_{i,k}-w_{i,k}) \notag \\
		&+c_2*r_2*({Gbest}_{i,k}-w_{i,k}),\\
		w_{i,k}=&Ve_{i,k}+w_{i,k},
	\end{align}
\end{subequations}
\noindent where $Ve_{i,k}$ is the velocity of MOPSO. $w_{i,k}\in\boldsymbol{w}_k$, $i=1, 2, \ldots, N_{\rm{BS}}$, and $k=1, 2, \ldots, K$. ${Pbest}_{i,k}$ and ${Gbest}_{i,k}$ correspond to the special dimension $\{i,k\}$ of individual historical optimal solution and obtained optimal solution, respectively. $r_1$ and $r_2$ are two random numbers generated between $(0, 1)$, and $\epsilon$ is the inertia factor. Moreover, $c_1$ and $c_2$ are learning factors, which can refer to \cite{DBLP:journals/tec/CoelloPL04}. However, using Eq. (\ref{comupdate}) may lead to the constraint violation, i.e., Eq. (\ref{En_eff_com}f). Thus, we propose a beamforming vector normalization operator shown in Algorithm \ref{Normalization}, in which $r_3$ is also a random number generated between $(0, 1)$.
\begin{algorithm}[tbp]
	\caption{Beamforming vector normalization operator}
	\label{Normalization}
	\textbf{Input:} $\boldsymbol{w}$ and $P_{\max}$.\\
	\For{$pop$= $1$ to $Pop$}{
	\eIf {$\boldsymbol{w}^H \boldsymbol{w} \leq P_{\max}$}{
	Do nothing;\\
}(\tcc*[f]{Eq. (\ref{En_eff_com}f) is not satisfied})
{
	$\boldsymbol{w}= \frac{\sqrt{P_{\max}}*\boldsymbol{w}}{\left\|\boldsymbol{w}\right\|*(1+r_3)}$;
}
}
\end{algorithm}

\subsection{Algorithm Analysis}

\par In this section, the complexity, convergence and drawbacks of the proposed algorithm are analyzed.
\subsubsection{Computation Complexity}
\label{Computation_complexity}

\par The computation complexity mainly depends on the computations of the objective functions and sorting the solutions in each objective function. Let $Obj$ represent the number of objective functions, which is $3$ in this work. Referred to \cite{DBLP:journals/tec/Jensen03a}, the computation complexity of the conventional NSGA-II is $O(Obj\cdot {Pop}^2\cdot G_{\max})$, when the dimension of decision variables is ignored. Moreover, \cite{DBLP:journals/tec/Jensen03a} assumes that the time to calculate each objective function is equal, while it is inapplicable for this work. Considering the dimensional changes, the computation complexity is $O({Pop}^2\cdot G_{\max}\cdot(M\cdot N_{\rm{RIS}}+N_{\rm{BS}}+3M))$, where $M$, $N_{\rm{RIS}}$ and $N_{\rm{BS}}$ are the number of UAV-RISs, RIS elements and the antennas of the BS, respectively. According to the proposed INSGA-II-CDC, the number of newly generated offspring is $4Pop$, and hence the computation complexity of INSGA-II-CDC is $O(16{Pop}^2\cdot G_{\max}\cdot(M\cdot N_{\rm{RIS}}+N_{\rm{BS}}+3M))$. When $Pop$ is sufficiently large, the computation complexity of INSGA-II-CDC is the same as the conventional NSGA-II.
\subsubsection{Convergence}
\par The formulated EEComm-MOF with three optimization objectives is a multi-objective optimization problem, which is inherently more intricate than the single-objective optimization problems because of the trade-offs and conflicts among optimization objectives. Faced with a multi-objective optimization problem with Pareto dominance, a common situation is that a solution is better on the first objective but worse on the second objective. Under these circumstances, it is hard to say whether this solution is better or worse. Thus, it is difficult to ensure that all optimization objectives converge simultaneously. In other words, it is challenging to derive the convergence of INSGA-II-CDC theoretically. Instead, we give the simulation results to verify the convergence, which is shown in Section \ref{Convergence and optimality verification}.

\subsubsection{Drawbacks}
\par As mentioned above, INSGA-II-CDC is based on the framework of evolutionary multi-objective optimization algorithm, hence it has the drawbacks of evolutionary multi-objective optimization algorithms. Specifically, these algorithms cannot guarantee to obtain the global optimal solution when the problem is complex. Moreover, the optimality of these algorithms are difficult to be analyzed theoretically. However, they are still feasible methods to solve NP-hard optimization problems for the practical scenarios, since these problems always contains a large number of decision variables, and using evolutionary multi-objective optimization algorithms can obtain an acceptable solution in polynomial time.

\section{Simulation Results}
\label{sec:Simulation Results}
\par In this section, simulation results are provided based on Matlab to illustrate the effectiveness of the proposed INSGA-II-CDC. First, the main parameter settings of the considered scenario are given. Then, we exploit INSGA-II-CDC and several benchmarks to solve the formulated EEComm-MOF. Afterward, we show the optimization results and evaluate the algorithm performance under different parameter settings. Finally, the implementability analysis is given.

\subsection{Parameter Settings}
\par The horizontal simulation scenario is set as $200$ m $\times 200$ m, and the maximum and minimum heights of the UAV-RISs are set as $200$ m and $50$ m, respectively. The parameters about UAV can refer to \cite{DBLP:journals/twc/ZengXZ19}. Other key parameters can be found in Table \ref{Para}.
\begin{table}[tbp]
	\setlength{\abovedisplayskip}{1pt}
	\setlength{\belowdisplayskip}{1pt}
	\setlength{\abovecaptionskip}{1pt}
	\centering
    \normalsize
	\caption{Key simulation parameters}
	\label{Para}
	\resizebox{\linewidth}{!}{
	\begin{tabular}{lll}
		\toprule
		\makebox[0.08\textwidth][l]{\textbf{Notation}} & \makebox[0.05\textwidth][l]{\textbf{Meaning}} & \makebox[0.04\textwidth][l]{\textbf{Value}}  \\ \midrule
		$B$      & Bandwidth      & 1 MHz  \\
		$P_{\max}$ & Maximum transmit power of BS & $50$ dBm \\
		$P_{\mathrm{BS}}$ & Circuit power of BS & $39$ dBm \\
		$\sigma^2$  &  Noise power  & $-104$ dBm \\
		$M_{\rm{UR}}$ & The weight of UAV-RIS    & $2$ kg  \\
		$P_k$  & Circuit power of each GU   & $10$ dBm  \\
		$P_{\mathrm{R}}$    & Circuit power of each RIS element  & $10$ dBm  \\
		$Q$    & Data size  & $10$ Mb \\
		$M$		& Number of UAV-RISs     & $\{2, 4, 6, 8\}$  \\
		$K$		& Number of GUs     & $\{5, 10\}$  \\
		$c$		& Number of bits for quantizing the phase shift levels     & $3$  \\
		$N_{\rm{BS}}$  &Number of BS antennas &   $32$  \\
		$N_{\rm{r}}$  & Number of RIS elements along $x$-axis & $8$   \\
		$N_{\rm{c}}$  & Number of RIS elements along $y$-axis & $8$   \\
		$\mathbf{q}_{\rm{B}}$  & Location of BS  & $[0, 0, 0]$  \\
		$\beta_0$   &Reference channel coefficient & $-30$ dB  \\
		$\mathcal{A}$ & Rician factor & $20$ dB \\
		$\mu$     & Power amplifier efficiency of BS       & $0.8$ \\ 
		$Pop$	& Population size       & $50$ \\ 
		$G_{\max}$	& Maximum iteration       & $200$ \\ 
		\bottomrule
		
	\end{tabular}
}
\end{table}

\begin{figure*}[h]
	\centering{\includegraphics[width=6in]{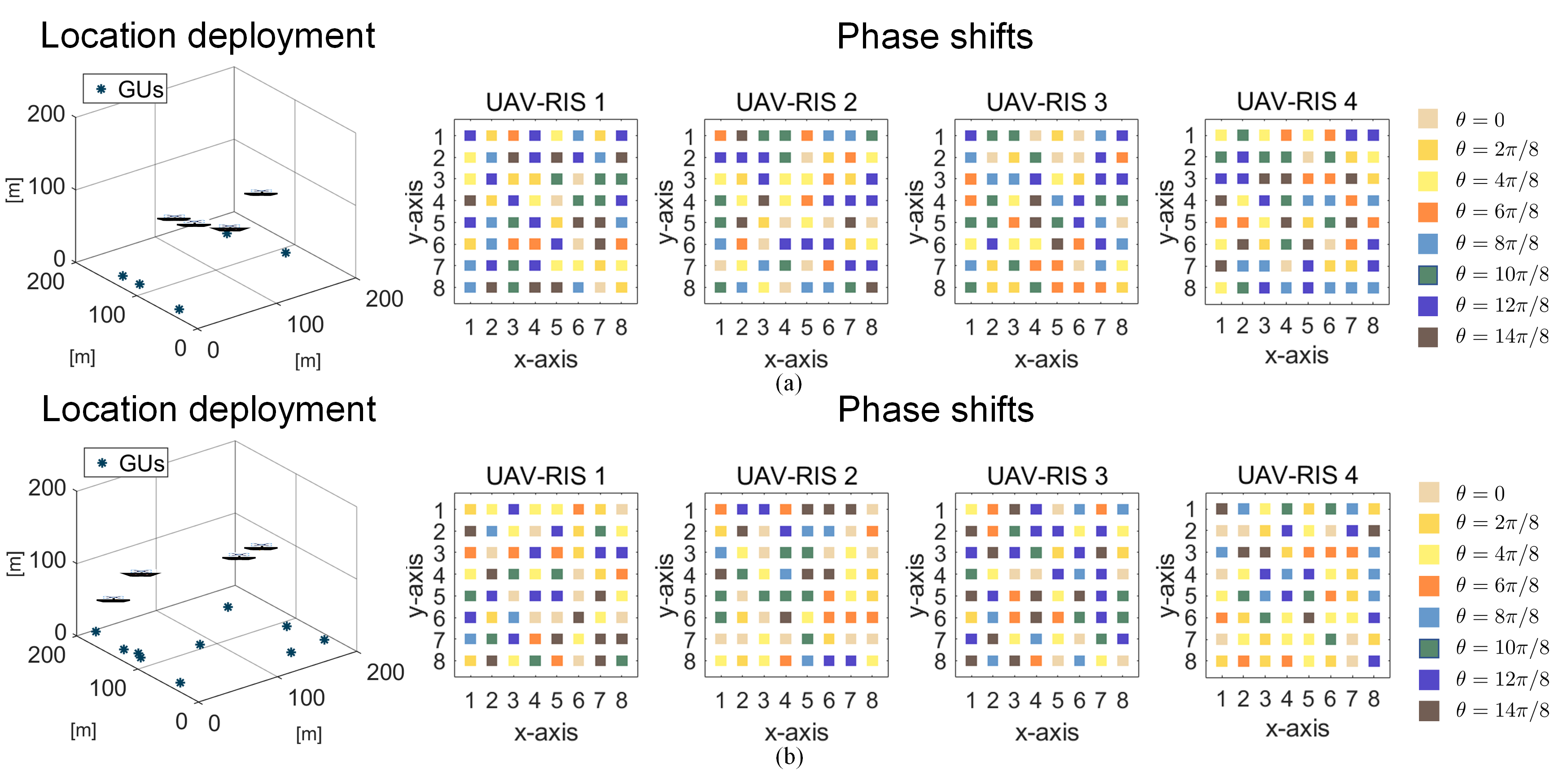}}
	\caption{Location deployment and phase shifts obtained by the proposed INSGA-II-CDC when $M=4$. (a) $5$ GUs. (b) $10$ GUs.}
	\label{Dply_Phase_Shift}
\end{figure*}
\begin{figure*}[h]
	\centering{\includegraphics[width=5.5in]{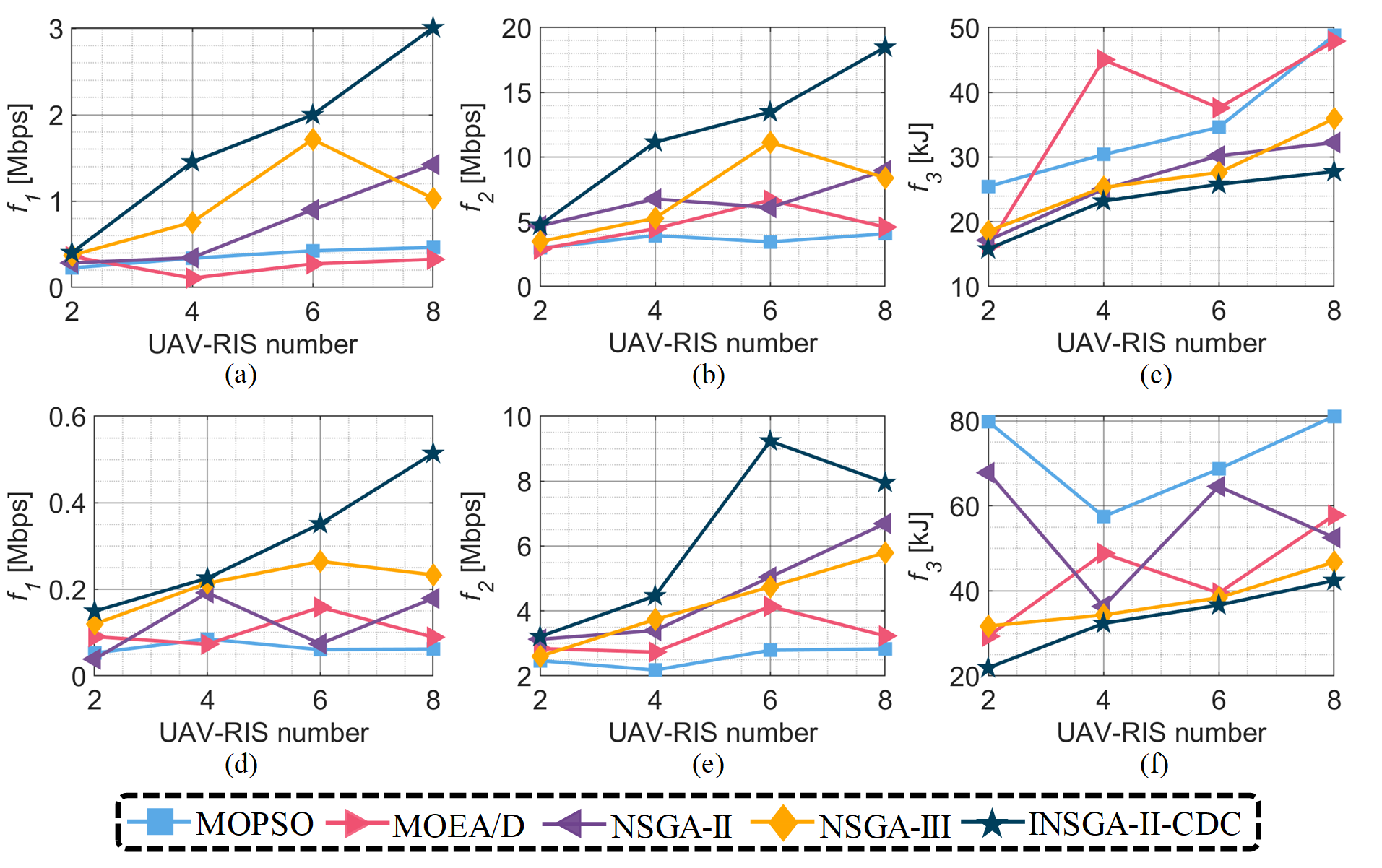}}
	\caption{Value of the objective functions versus UAV-RIS number. (a) $f_1$ vs UAV-RIS number for $5$ GUs. (b) $f_2$ vs UAV-RIS number for $5$ GUs. (c) $f_3$ vs UAV-RIS number for $5$ GUs. (d) $f_1$ vs UAV-RIS number for $10$ GUs. (e) $f_2$ vs UAV-RIS number for $10$ GUs. (f) $f_3$ vs UAV-RIS number for $10$ GUs.}
	\label{fvsUAVnum}
\end{figure*}

\begin{figure*}[h]
	\centering{\includegraphics[width=4.5in]{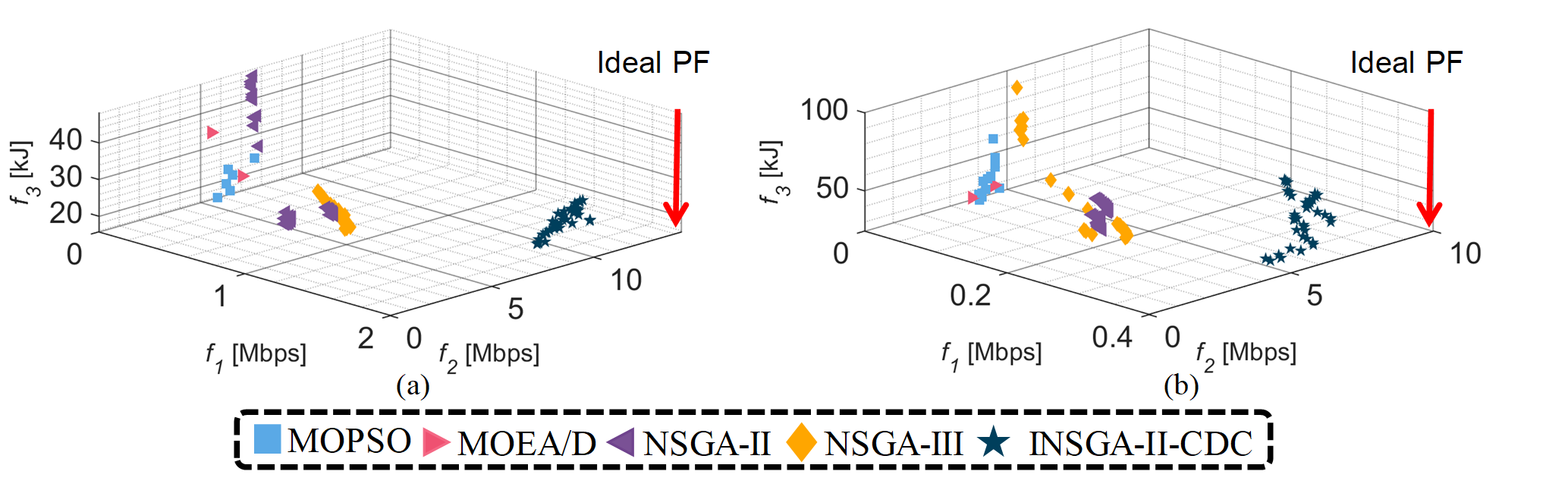}}
	\caption{\textcolor{black}{Pareto solution distributions obtained by the proposed INSGA-II-CDC and other baseline algorithms when $M=4$. (a) $5$ GUs. (b) $10$ GUs.}}
		\label{Pareto_front}
\end{figure*}

\begin{figure*}[h]
	\centering{\includegraphics[width=4.5in]{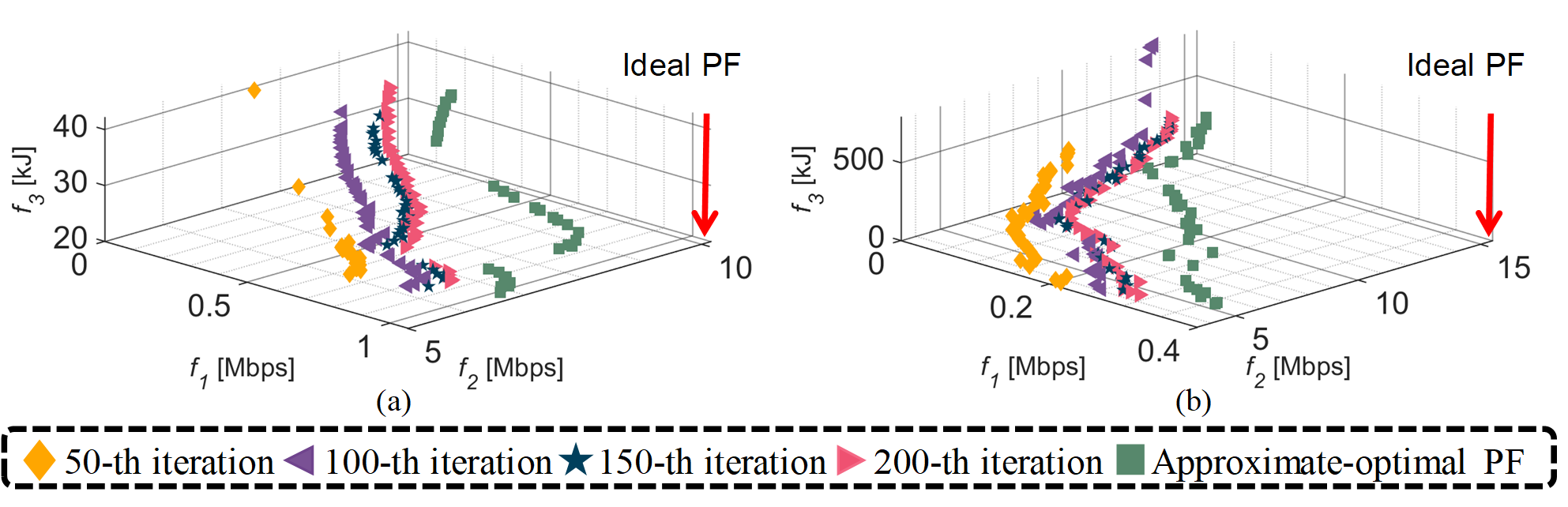}}
	\caption{Advanced progress of PF and gap with the approximate-optimal
		PF obtained by INSGA-II-CDC. (a) $5$ GUs. (b) $10$ GUs.}
	\label{Convergence}
\end{figure*}

\subsection{Benchmarks}
\par Several benchmark strategies are introduced in this paper. 

\begin{figure*}[t]
\centering{\includegraphics[width=5.5in]{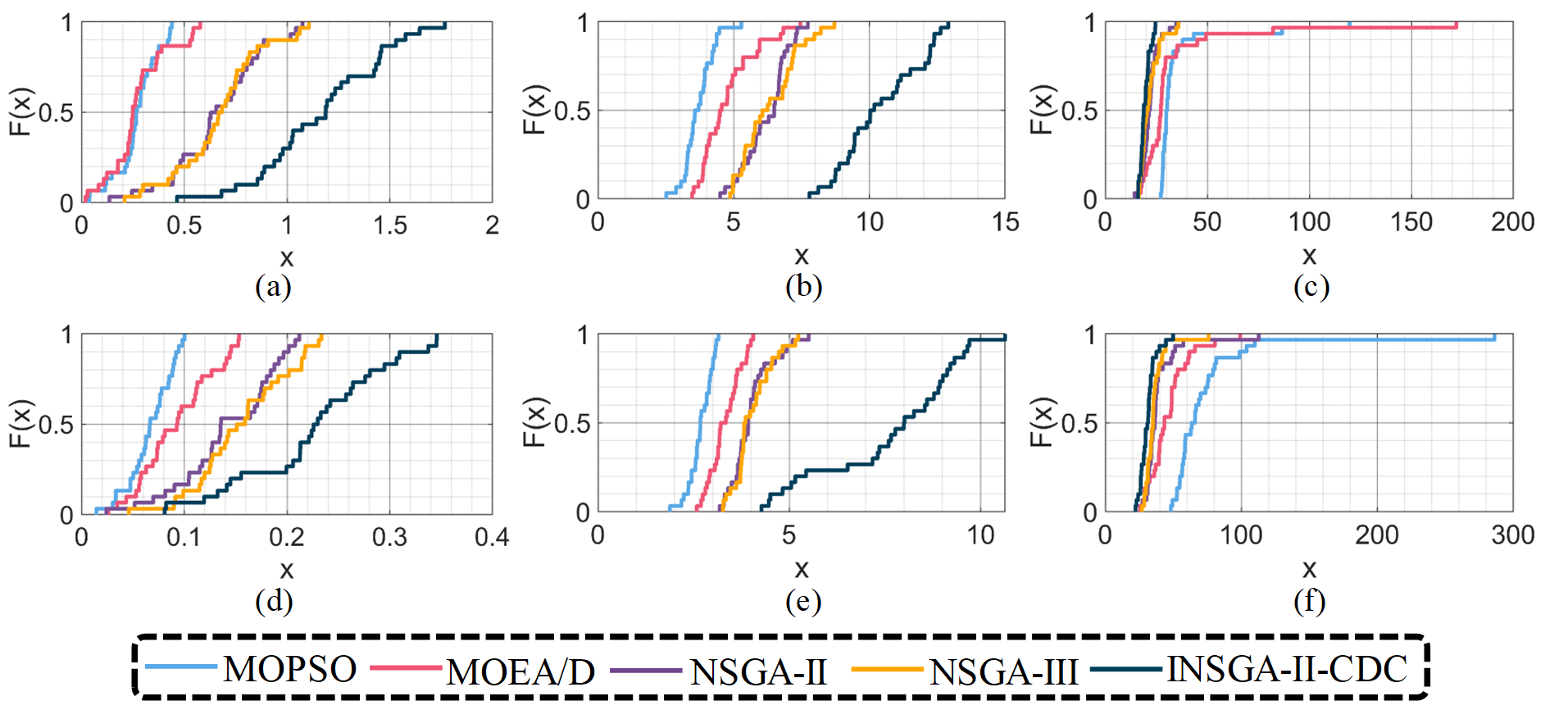}}
	\caption{CDFs obtained by the proposed INSGA-II-CDC and other baseline algorithms when $M=4$. (a) $f_1$ for $5$ GUs. (b) $f_2$ for $5$ GUs. (c) $f_3$ for $5$ GUs. (d) $f_1$ for $10$ GUs. (e) $f_2$ for $10$ GUs. (f) $f_3$ for $10$ GUs.}
\label{CDF_f}
\end{figure*}
\begin{itemize}
\item \textbf{Random Deployment (RD)}: In RD strategy, the UAV-RIS location deployment, phase shifts, and beamforming vector are randomly generated.
\item \textbf{Uniform Deployment (UD)}: In UD strategy, the horizontal UAV-RIS locations are deployed uniformly, while the height of UAV-RISs is set as $\frac{Z_{\max}+Z_{\min}}{2}$. The phase shifts of all elements is set as $\pi$, while the beamforming vector is randomly generated.
\item \textbf{Discrete Fourier Transform Design (DFT-Design)}: In DFT-design, the UAV-RIS location deployment and beamforming vector are randomly generated, while the phase shifts will be determined by discrete Fourier transform.
\item \textbf{Classic Discrete Phase Shift Design (CDPS-Design)}: In CDPS-design, the UAV-RIS location deployment and beamforming vector are also randomly generated, while the phase shifts will be iteratively optimized dimension by dimension, and the detailed process can be found in \cite{DBLP:journals/tcom/WuZ20}.
\item \textbf{Baseline Algorithms}: Several evolutionary multi-objective optimization algorithms, which are MOPSO \cite{DBLP:journals/tec/CoelloPL04}, multi-objective evolutionary algorithm based on decomposition (MOEA/D) \cite{DBLP:journals/tcyb/ZhangGLSZT22}, NSGA-II \cite{DBLP:journals/tec/DebAPM02}, and non-dominated sorting genetic algorithm-III (NSGA-III) \cite{DBLP:journals/tpds/ZhouRXF23} are employed as baseline algorithms. In addition, Algorithm \ref{LargePhase} is used to handle with the discrete phase shifts, while Algorithm \ref{Normalization} is adopted to satisfy the transmit power constraint for all baseline algorithms.

\end{itemize}

\begin{table}[h]
	\begin{center}
		\scriptsize
		\caption{Numerical statistical results obtained by different benchmarks when $M=4$ for $5$ GUs}
		\setlength{\tabcolsep}{0.6mm}
		{\begin{tabular}{|c|c|c|c|c|c|c|}\hline	
				&Benchmarks &{Mean} &{Std.} &{Max} &{Min} &{Improvement}\\
				\hline
				\multirow{9}{*}{\textbf {$f_1$ [Mbps]}}
				&\textbf {RD}  		&$0.06$ &$0.06$ &$0.28$ &$0.00$ &$-$\\
				&\textbf {UD}  		&$0.03$ &$\bf{0.04}$ &$0.23$ &$0.00$ &$-$\\
				&\textbf {DFT-Design}  		&$0.25$ &$0.09$ &$0.47$ &$0.12$ &$-$\\
				&\textbf {CDPS-Design}  	&$0.54$ &$0.17$ &$0.89$ &$0.26$ &$-$\\
				&\textbf{MOPSO}  	&$0.27$ &$0.10$ &$0.44$ &$0.04$ &$-$\\
				&\textbf{MOEA/D}  	&$0.27$ &$0.14$ &$0.58$ &$0.02$ &$-$\\
				&\textbf{NSGA-II}  	&$0.66$ &$0.22$ &$1.07$ &$0.14$ &$-$\\
				&\textbf{NSGA-III}  &$0.67$ &$0.22$ &$1.11$ &$0.21$ &$-$\\
				&\textbf{INSGA-II-CDC} 	&$\bf{1.17}$ &$0.30$ &$\bf{1.77}$ &$\bf{0.46}$ &$74.62\%$\\
				\hline
				\multirow{9}{*}{\textbf {$f_2$ [Mbps]}}
				&\textbf {RD}  		&$1.85$ &$0.58$ &$3.38$ &$0.94$ &$-$\\
				&\textbf {UD}  		&$0.83$ &$\bf{0.41}$ &$1.90$ &$0.35$ &$-$\\
				&\textbf {DFT-Design}  		&$2.27$ &$0.62$ &$4.23$ &$1.16$ &$-$\\
				&\textbf {CDPS-Design}  		&$3.49$ &$0.75$ &$5.60$ &$2.35$ &$-$\\
				&\textbf{MOPSO}  	&$3.71$ &$0.56$ &$5.28$ &$2.51$ &$-$\\
				&\textbf{MOEA/D}  	&$4.78$ &${1.02}$ &$7.45$ &$3.46$ &$-$\\
				&\textbf{NSGA-II}  	&$6.21$ &$0.83$ &$7.73$ &$4.50$ &$-$\\
				&\textbf{NSGA-III}  &$6.33$ &$1.05$ &$8.70$ &$4.86$ &$-$\\
				&\textbf{INSGA-II-CDC} 	&$\bf{10.41}$ &$1.49$ &$\bf{12.90}$ &$\bf{7.79}$ &$64.45\%$\\
				\hline
				\multirow{9}{*}{\textbf {$f_3$ [kJ]}}
				&\textbf {RD}  		&$159.77$ &$198.24$ &$922.55$ &$27.41$ &$-$\\
				&\textbf {UD}  		&$548.50$ &$2069.38$ &$11479.60$ &$27.20$ &$-$\\
				&\textbf {DFT-Design}  		&$31.20$ &$3.68$ &$36.31$ &$23.14$ &$-$\\
				&\textbf {CDPS-Design}  		&$25.01$ &$3.58$ &$37.14$ &$19.11$ &$-$\\
				&\textbf{MOPSO}  	&$35.61$ &$19.16$ &$119.64$ &$27.18$ &$-$\\
				&\textbf{MOEA/D}  	&$33.62$ &$28.90$ &$171.93$ &$16.20$ &$-$\\
				&\textbf{NSGA-II}   &$21.97$ &$4.23$ &$34.76$ &$\bf{14.29}$ &$-$\\
				&\textbf{NSGA-III}  &$21.80$ &$4.91$ &$35.83$ &$16.36$ &$-$\\
				&\textbf{INSGA-II-CDC} 	&$\bf{19.50}$ &$\bf{2.40}$ &$\bf{24.60}$ &$16.03$ &$10.55\%$\\
				\hline
		\end{tabular}}
		\label{Num_small}
	\end{center}
\end{table}

\begin{table}[t]
	\begin{center}
		\scriptsize
            \caption{Numerical statistical results obtained by different benchmarks when $M=4$ for $10$ GUs}
		\setlength{\tabcolsep}{0.6mm}
		{\begin{tabular}{|c|c|c|c|c|c|c|}\hline	
				&Benchmarks &{Mean} &{Std.} &{Max} &{Min} &{Improvement}\\
				\hline
				\multirow{9}{*}{\textbf {$f_1$ [Mbps]}}
				&\textbf {RD}  	&$0.02$ &$0.02$ &$0.08$ &$0.00$ &$-$\\
				&\textbf {UD}  		&$0.01$ &$\bf{0.01}$ &$0.04$ &$0.00$ &$-$\\
				&\textbf {DFT-Design}  		&$0.07$ &$0.01$ &$0.10$ &$0.04$ &$-$\\
				&\textbf {CDPS-Design}  	&$0.14$ &$0.06$ &$0.28$ &$0.02$ &$-$\\
				&\textbf{MOPSO}  	&$0.07$ &$0.02$ &$0.10$ &$0.01$ &$-$\\
				&\textbf{MOEA/D}  	&$0.09$ &$0.04$ &$0.15$ &$0.03$ &$-$\\
				&\textbf{NSGA-II}  	&$0.14$ &$0.05$ &$0.21$ &$0.02$ &$-$\\
				&\textbf{NSGA-III}  &$0.16$ &$0.05$ &$0.23$ &$0.05$ &$-$\\
				&\textbf{INSGA-II-CDC} 	&$\bf{0.23}$ &$0.07$ &$\bf{0.35}$ &$\bf{0.09}$ &$43.75\%$\\
				\hline
				\multirow{9}{*}{\textbf {$f_2$ [Mbps]}}
				&\textbf {RD}  	&$1.57$ &$0.36$ &$2.26$ &$0.93$ &$-$\\
				&\textbf {UD}  		&$0.87$ &$\bf{0.29}$ &$1.50$ &$0.35$ &$-$\\
				&\textbf {DFT-Design}  		&$1.68$ &$0.32$ &$2.26$ &$1.15$ &$-$\\
				&\textbf {CDPS-Design}  		&$2.35$ &$0.58$ &$3.75$ &$1.35$ &$-$\\
				&\textbf{MOPSO}  	&$2.69$ &$0.30$ &$3.15$ &$1.88$ &$-$\\
				&\textbf{MOEA/D}  	&$3.32$ &$0.41$ &$4.05$ &$2.59$ &$-$\\
				&\textbf{NSGA-II}  	&$4.00$ &$0.55$ &$5.51$ &$3.18$ &$-$\\
				&\textbf{NSGA-III}  &$4.03$ &$0.51$ &$5.23$ &$3.27$ &$-$\\
				&\textbf{INSGA-II-CDC} 	&$\bf{7.64}$ &$1.82$ &$\bf{10.64}$ &$\bf{4.27}$ &$89.57\%$\\
				\hline
				\multirow{9}{*}{\textbf {$f_3$ [kJ]}}
				&\textbf {RD}  	&$381.75$ &$587.05$ &$2569.60$ &$43.72$ &$-$\\
				&\textbf {UD}  		&$1877.57$ &$5521.91$ &$30434.78$ &$72.08$ &$-$\\
				&\textbf {DFT-Design}  		&$56.27$ &$9.00$ &$81.46$ &$45.99$ &$-$\\
				&\textbf {CDPS-Design}  		&$43.43$ &$23.94$ &$155.78$ &$28.52$ &$-$\\
				&\textbf{MOPSO}  	&$74.82$ &$42.79$ &$286.04$ &$48.13$ &$-$\\
				&\textbf{MOEA/D}  	&$46.70$ &$15.89$ &$99.10$ &$25.36$ &$-$\\
				&\textbf{NSGA-II}   &$39.48$ &$15.47$ &$112.81$ &$26.78$ &$-$\\
				&\textbf{NSGA-III}  &$36.24$ &$8.72$ &$75.93$ &$26.99$ &$-$\\
				&\textbf{INSGA-II-CDC} 	&$\bf{31.31}$ &$\bf{6.00}$ &$\bf{49.81}$ &$\bf{22.12}$ &$13.60\%$\\
				\hline
		\end{tabular}}
		\label{Num_large}
	\end{center}
\end{table}

\subsection{Optimization Results}
\par In this section, we show the visualization results and test the effectiveness of INSGA-II-CDC for different number of UAV-RISs.

\subsubsection{Visualization Results}
\par Figs. \ref{Dply_Phase_Shift}(a) and \ref{Dply_Phase_Shift}(b) show the location deployment and phase shifts obtained by the proposed INSGA-II-CDC when $M=4$ for $5$ GUs and $10$ GUs, respectively. As shown in the figures for phase shifts, different colors mean different phase shifts. For the ease of presentation, the Pareto solution distributions obtained by the proposed INSGA-II-CDC and other baseline algorithms when $M=4$ are shown in Figs. \ref{Pareto_front}(a) and \ref{Pareto_front}(b), for $5$ GUs and $10$ GUs, respectively. Apparently, the solutions obtained by INSGA-II-CDC are much closer to the ideal PF, i.e., the optimal values of the three optimization objectives. The reason can be that the continuous solution processing mechanism enhances the population diversity, hence improving the search efficiency, while the discrete and complex solution processing mechanisms improve the quality of solutions iteratively. Thus, the proposed INSGA-II-CDC is more suitable to solve the formulated EEComm-MOF.

\subsubsection{Effectiveness of INSGA-II-CDC for Different Values of $M$}
\par Moreover, we test the results of different algorithms under different values of $M$ for $5$ GUs which can be seen in Figs. \ref{fvsUAVnum}(a)-\ref{fvsUAVnum}(c), while the similar results for $10$ GUs are shown in Figs. \ref{fvsUAVnum}(d)-\ref{fvsUAVnum}(f). To ensure the fairness, we select the median of the obtained PF to plot these figures. As can be seen, the proposed INSGA-II-CDC performs better than other baseline algorithms. Overall, as the number of UAV-RISs increases, INSGA-II-CDC achieves improvements in both the minimum available rate over all GUs and the total available rate of GUs. However, the total energy consumption of the system also increases. This can be attributed to the fact that more UAV-RISs result in stronger reflected signals but also require additional energy. Interestingly, in Fig. \ref{fvsUAVnum}(e), the value of $f_2$ obtained by INSGA-II-CDC decreases when transitioning from $M=6$ to $M=8$. The reason for this phenomenon is that the case with $M=8$ is more complex due to more decision variables. Moreover, there are inherent trade-offs among the three optimization objectives. As a result, the search process encounters greater difficulties, and INSGA-II-CDC may discard different solutions in the elitist filter, as described in Algorithm \ref{INSGA-II-CDC}. This discrepancy may cause a slight deviation in the direction of the obtained PF, and the fluctuation obtained by other baseline algorithms in Fig. \ref{fvsUAVnum} also follows this reason.

\subsection{Algorithm Performance Evaluation}
\par In this section, we first verify the convergence and the optimality. Then, the stability of INSGA-II-CDC and effectiveness of improved mechanisms are evaluated. Finally, the CPU running time of different baseline algorithms is given.

\subsubsection{Convergence and Optimality Verification}
\label{Convergence and optimality verification}

\par As mentioned, proving the convergence of INSGA-II-CDC theoretically is challenging. Thus, it is reasonable to use the advanced process of PF to verify the convergence of the algorithm by introducing the concept of Pareto dominance \cite{DBLP:journals/twc/MehariPCDVPJMDM16}. Specifically, Fig. \ref{Convergence}(a) shows the advanced process of PF in different iterations for the case of $5$ GUs, while Fig. \ref{Convergence}(b) shows the corresponding results for the case of $10$ GUs. It can be seen from Fig. \ref{Convergence}(a) that the obtained PF on $200$-th iteration is better than that on $150$-th iteration, which means that the obtained solution qualities are getting better with more iterations, while the improvement in solution qualities decreases gradually. Thus, the algorithm gradually converges. Moreover, the similar result can also be found in Fig. \ref{Convergence}(b), while the magnitude of the advance on PF from $50$-th iteration to $100$-th iteration is less obvious than the case of $5$ GUs. The reason may be that the solution space of the case of $10$ GUs is more complex, which means that the algorithm needs more searches to find a better PF. 

\par Moreover, since the three optimization objectives contain trade-offs, it is difficult to find the optimal values for each optimization objective simultaneously. Thus, finding the approximate-optimal solution for the problem may be more reasonable, which is also a common way for the multi-objective optimization optimization problems in wireless systems and experimentations \cite{DBLP:journals/twc/MehariPCDVPJMDM16}. To analyze the approximate gap in a feasible way, we first set the number of iterations of INSGA-II-CDC to $1000$, which is a much larger value than the normal simulations. Then, the obtained optimization result is regarded as the approximate-optimal PF, and we compare the gap between the result of the normal simulations and the approximate-optimal PF, and the corresponding results are shown in Figs. \ref{Convergence}(a) and \ref{Convergence}(b). As can be seen, the PF obtained by $200$-th iteration is close to the approximate-optimal PF, which mean that the optimality of the algorithm can be verified.

\subsubsection{Stability of INSGA-II-CDC}
\par To verify the stability of INSGA-II-CDC, we visualize the cumulative distribution functions (CDFs) obtained by INSGA-II-CDC and the baseline algorithms when $M=4$ in Fig. \ref{CDF_f}. Obviously, the optimization results obtained by INSGA-II-CDC dominate other baseline algorithms, which means that INSGA-II-CDC can obtain a better stability whether for $5$ GUs or $10$ GUs. The abovementioned results exactly correspond to the numerical statistical results in Tables \ref{Num_small} and \ref{Num_large}, respectively, where ``Mean'', ``Std.'', ``Max'' and ``Min'' represent the mean value, standard deviation, maximum value and minimum value of $30$ independent trials, respectively \footnote{According to the central limit theorem \cite{DBLP:journals/joi/Antonoyiannakis18}, taking $30$ independent trials is generally accepted by statisticians and researchers to assume the distribution of sample mean approximated to normal distribution.}. Moreover, ``Improvement'' refers to the improvement ratio of INSGA-II-CDC on the corresponding objectives compared with the suboptimal benchmarks. Specifically, for a cellular network with $5$ GUs, we can increase the minimum available rate by $74.62\%$, while increasing the total available rate by $64.45\%$, when the energy consumption is saved by $10.55\%$. Similarly, for a cellular network with $10$ GUs, we can increase the minimum available rate by $43.75\%$, while increasing the total available rate by $89.57\%$, when the energy consumption is saved by $13.60\%$.

\subsubsection{Effectiveness of Improved Mechanisms}

\par \textcolor{black}{In this section, we conduct tests to verify the effectiveness of the introduced improved mechanisms. Specifically, the improved NSGA-II with a continuous solution processing mechanism (INSGA-II-C1), the improved NSGA-II with a discrete solution processing mechanism (INSGA-II-D), and the improved NSGA-II with a complex solution processing mechanism (INSGA-II-C2) are used to solve EEComm-MOF, respectively. Tables \ref{Improved_effective_small} and \ref{Improved_effective_large} show the optimization results obtained by NSGA-II with part of improved mechanisms. As can be seen, INSGA-II-C1, INSGA-II-D, and INSGA-II-C2 obtain better optimization results on most of the optimization objectives, respectively, which means that each of improved mechanisms is necessary, Moreover, the proposed INSGA-II-CDC which combines all the advantages of these improved mechanisms can improve the results of all different optimization objectives, and hence it is effective. }                      

\begin{table}[htb]
	\begin{center}
		\scriptsize
		\caption{\textcolor{black}{Numerical statistical results obtained by NSGA-II with part of improved mechanisms when $M=4$ for $5$ GUs}}
		\setlength{\tabcolsep}{1.5mm}
		{\begin{tabular}{|l|l|l|l|l|l|}\hline	
				&Algorithm &{Mean} &{Std.} &{Max} &{Min} \\
				\hline
				\multirow{5}{*}{\textbf{$f_1$ [Mbps]}}				
			&\textbf{NSGA-II}  	 
                &$0.66$ &$0.22$ &$1.07$ &$0.14$ \\
                &\textbf{INSGA-II-C1} &$0.67$ &$0.23$ &$1.09$ &$0.12$ \\
                &\textbf{INSGA-II-D} &$0.71$ &$\bf{0.21}$ &$1.08$ &$0.29$ \\
                &\textbf{INSGA-II-C2} &$0.61$ &$0.25$ &$1.02$ &$0.10$ \\
			&\textbf{INSGA-II-CDC} 	 
                &$\bf{1.17}$ &$0.30$ &$\bf{1.77}$ &$\bf{0.46}$ \\
				\hline
                    \multirow{5}{*}{\textbf{$f_2$ [Mbps]}}
    			&\textbf{NSGA-II}  	 
                &$6.21$ &$0.83$ &$7.73$ &$4.50$ \\
                &\textbf{INSGA-II-C1} &$6.50$ &$\bf{0.66}$ &$7.76$ &$5.02$ \\
                &\textbf{INSGA-II-D} &$6.80$ &$1.08$ &$9.00$ &$5.11$ \\
                &\textbf{INSGA-II-C2} &$7.06$ &$0.95$ &$9.24$ &$5.48$ \\
			&\textbf{INSGA-II-CDC} 	 
                &$\bf{10.41}$ &$1.49$ &$\bf{12.90}$ &$\bf{7.79}$ \\
				\hline
                \multirow{5}{*}{\textbf{$f_3$ [kJ]}}
                    &\textbf{NSGA-II}  	 
                &$21.97$ &$4.23$ &$34.76$ &$\bf{14.29}$ \\
                &\textbf{INSGA-II-C1} &$22.88$ &$4.89$ &$39.85$ &$17.03$ \\
                &\textbf{INSGA-II-D} &$21.58$ &$3.02$ &$30.44$ &$15.07$ \\
                &\textbf{INSGA-II-C2} &$21.39$ &$4.92$ &$37.81$ &$16.73$ \\
			&\textbf{INSGA-II-CDC} 	 
                &$\bf{19.50}$ &$\bf{2.40}$ &$\bf{24.60}$ &$16.03$ \\
				\hline
		\end{tabular}}
		\label{Improved_effective_small}
	\end{center}
\end{table}
\begin{table}[htb]
	\begin{center}
		\scriptsize
		\caption{\textcolor{black}{Numerical statistical results obtained by NSGA-II with part of improved mechanisms when $M=4$ for $10$ GUs}}
		\setlength{\tabcolsep}{1.5mm}
		{\begin{tabular}{|l|l|l|l|l|l|}\hline	
				&Algorithm &{Mean} &{Std.} &{Max} &{Min} \\
				\hline
				\multirow{5}{*}{\textbf{$f_1$ [Mbps]}}				
			&\textbf{NSGA-II}  	 
                &$0.14$ &$\bf{0.05}$ &$0.21$ &$0.02$ \\
                &\textbf{INSGA-II-C1} &$0.15$ &$\bf{0.05}$ &$0.22$ &$0.06$ \\
                &\textbf{INSGA-II-D} &$0.14$ &$\bf{0.05}$ &$0.24$ &$0.05$ \\
                &\textbf{INSGA-II-C2} &$0.15$ &$\bf{0.05}$ &$0.23$ &$0.07$ \\
			&\textbf{INSGA-II-CDC} 	 
                &$\bf{0.23}$ &$0.07$ &$\bf{0.34}$ &$\bf{0.08}$ \\
			\hline
                \multirow{5}{*}{\textbf{$f_2$ [Mbps]}}
    			&\textbf{NSGA-II}  	 
                &$4.00$ &$0.55$ &$5.51$ &$3.18$ \\
                &\textbf{INSGA-II-C1} &$4.12$ &$0.42$ &$4.96$ &$3.41$ \\
                &\textbf{INSGA-II-D} &$4.14$ &$4.33$ &$5.44$ &$3.38$ \\
                &\textbf{INSGA-II-C2} &$4.29$ &$\bf{0.35}$ &$4.92$ &$3.50$ \\
			&\textbf{INSGA-II-CDC} 	 
                &$\bf{7.64}$ &$1.82$ &$\bf{10.64}$ &$\bf{4.27}$ \\
				\hline
                \multirow{5}{*}{\textbf{$f_3$ [kJ]}}
                    &\textbf{NSGA-II}  	 
                &$39.48$ &$15.47$ &$112.80$ &$26.78$ \\
                &\textbf{INSGA-II-C1} &$36.94$ &$8.06$ &$55.31$ &$26.90$ \\
                &\textbf{INSGA-II-D} &$39.89$ &$11.16$ &$81.19$ &$27.86$ \\
                &\textbf{INSGA-II-C2} &$38.63$ &$8.24$ &$64.40$ &$28.89$ \\
			&\textbf{INSGA-II-CDC} 	 
                &$\bf{31.31}$ &$\bf{6.00}$ &$\bf{49.81}$ &$\bf{22.12}$ \\
				\hline
		\end{tabular}}
		\label{Improved_effective_large}
	\end{center}
\end{table}
\begin{table}[htb]
	\begin{center}
	\setlength{\abovedisplayskip}{1pt}
	\setlength{\belowdisplayskip}{1pt}
	\setlength{\abovecaptionskip}{5pt}
	\scriptsize 
	\tabcolsep=0.5mm
	\caption{Numerical statistical results of CPU running times obtained by INSGA-II-CDC and other baseline algorithms when $M=4$}
	\label{Time-NSGA-II-CDC}
		\begin{tabular}{|c|c|c|c|c|c|}\hline
			\textbf{Different $K$ value} & \textbf{MOPSO} &\textbf{MOEA/D}&\textbf{NSGA-II}&\textbf{NSGA-III} &\textbf{INSGA-II-CDC}\\\hline
			\textbf{$K=5$ [s]} & $\bm{27.30}$ &$29.40$ &$49.42$ &$50.52$ &$113.20$\\
			\textbf{$K=5$ [s]} & $\bm{29.40}$ &$53.33$ &$77.64$ &$77.79$ &$161.96$\\\hline
		\end{tabular}
	\end{center}
\end{table}
\subsubsection{CPU Running Time}
\par The numerical results of the CPU running times of INSGA-II-CDC and other baseline algorithms when $M=4$ are shown in Table \ref{Time-NSGA-II-CDC}. It can be seen from the table that INSGA-II-CDC takes longer CPU running time, since it takes additional calculations as the abovementioned analysis in Section \ref{Computation_complexity}. However, the gaps of CPU running times are not very large. Moreover, due to the fixed location of GUs, all UAV-RISs can be deployed and recalled synchronously once the energy is exhausted or the transmission task is completed. In other words, the algorithm can be run off-line, and the CPU running time is not a primary consideration. Thus, we may say that INSGA-II-CDC has the overall best performance for dealing with the formulated EEComm-MOF.

\subsection{Implementability Analysis}
\par
\begin{figure}[htb]
	\centering{\includegraphics[width=3.5in]{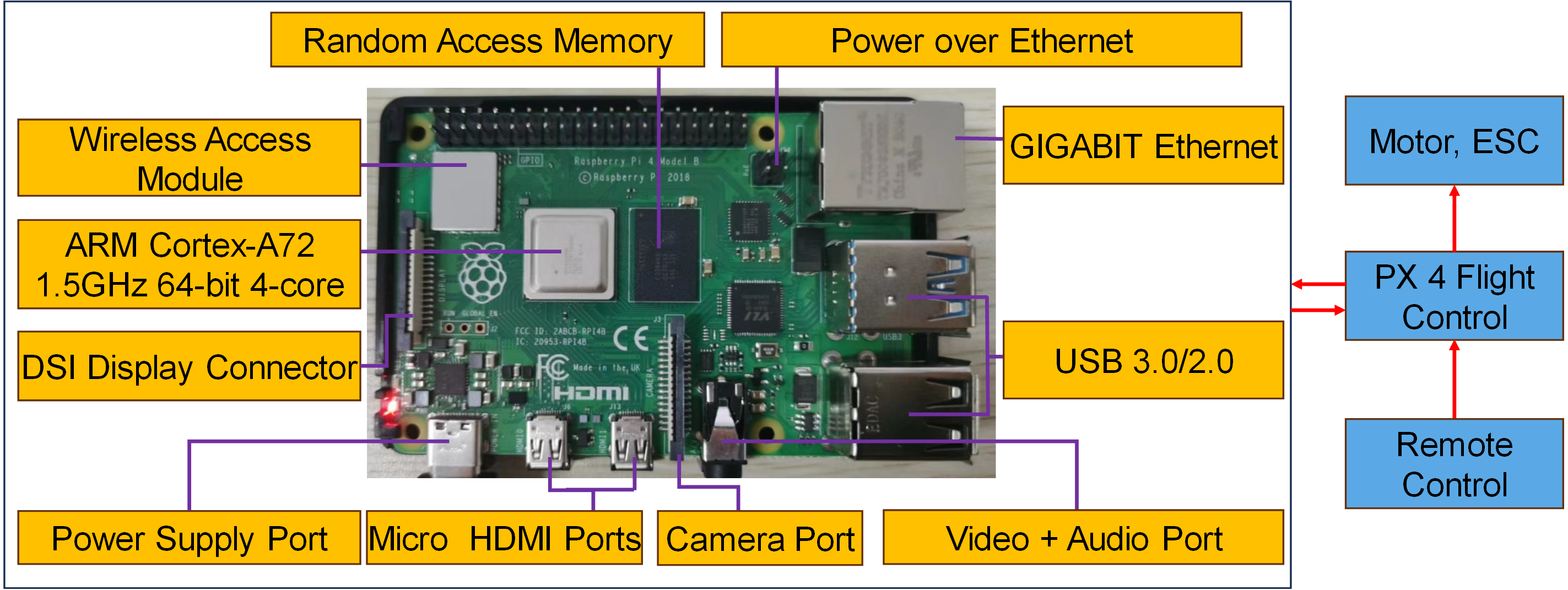}}
	\caption{Schematic diagram for the controller of UAV-RIS system based on Raspberry Pi.}
	\label{UAV-RIS experiments}
\end{figure}

 To verify the implementability of the system, we use the Raspberry Pi 4B to conduct experiments for the baseline algorithms. Generally, Raspberry Pi 4B platform is a processor commonly utilized in a practical UAV flight control system \cite{10012331}. Similar to a small-sized minicomputer, it can also transmit the commands to the RIS controller so as to control RIS phase shifts \cite{DBLP:journals/wcl/ZhouHJMD21}. The schematic diagram depicting the autonomous UAV-RIS system, based on the Raspberry Pi, is illustrated in Fig. \ref{UAV-RIS experiments}. Since the Raspberry Pi 4B platform is not compatible with Matlab, we translate the Matlab-programmed INSGA-II-CDC into Python. We neglect the computations of optimization objective values, as proxy models can effectively take their place in practical scenarios \cite{jeong2005efficient}. 

\par In addition, we implement other baseline algorithms, including MOPSO, MOEA/D, NSGA-II, and NSGA-III. Specifically, the execution times obtained by MOPSO, MOEA/D, NSGA-II, NSGA-III, and INSGA-II-CDC are $35.50$ s, $47.75$ s, $195.59$ s, $317.64$ s, and $779.67$ s, respectively. Despite the longer execution time of INSGA-II-CDC due to more calculations, it is still reasonable \cite{10278101}. The reasons can be summarized as follows: First, this work explores the simultaneous deployment of all UAV-RISs to assist the cellular network. To ensure seamless coverage, all UAV-RISs are recalled and a new batch is deployed when one UAV-RIS exhausts its energy. Thus, if the maximum hovering time of the UAV-RIS is greater than the computation time of the algorithm, the solution remains viable, since the algorithm can be run in advance when the last batch of UAV-RISs starts service. According to \cite{DBLP:journals/tits/MuntahaHJH21}, the UAV-RIS can hover for over $15$ minutes in dense urban environments, well surpassing the required computation time. Second, during actual deployment, the computation time is expected to be lower than the obtained results due to a lower code execution efficiency of Python compared to C/C++. Hence, deploying the C/C++ version of the algorithm would further enhance execution efficiency in real-world scenarios. Moreover, the computation power is also insignificant compared to the propulsion power of the UAV-RISs \cite{10278101}, which can be easily tackled.

\section{Conclusion}
\label{sec:Conclusion}

\par In this paper, a cooperative UAV-RISs-assisted cellular network is investigated, where multiple RISs are carried and enhanced by UAVs to serve multiple GUs simultaneously, thereby achieving 3D mobility and opportunistic deployment. Specifically, EEComm-MOF is formulated to jointly consider the beamforming vector of BS, the location deployment, and the discrete phase shifts of UAV-RISs so as to maximize the minimum available rate over all GUs, maximize the total available rate of all GUs, and minimize the total energy consumption of the system simultaneously, while satisfying the transmit power constraint of BS. Then, we propose an INSGA-II-CDC with several specific designs to solve the problem directly. Simulations results demonstrate that the proposed INSGA-II-CDC is better than other benchmarks under different parameter settings. Moreover, the performance of the INSGA-II-CDC in terms of the convergence and optimality, stability, effectiveness of improved mechanisms, and CPU running time is verified. In addition, the implementability of the proposed algorithm in practical system is evaluated. In the future work, we will consider the mobile GUs instead of fixed GUs, which means that the trajectories of UAV-RISs will be investigated.


%

\ifCLASSOPTIONcompsoc

\ifCLASSOPTIONcaptionsoff
  \newpage
\fi

\bibliographystyle{ieeetr}
\bibliography{myref}

\begin{thebibliography}{10}

\bibitem{DBLP:journals/wc/HuangHAZYZRD20}
C.~Huang, S.~Hu, G.~C. Alexandropoulos, A.~Zappone, C.~Yuen, R.~Zhang, M.~D.
  Renzo, and M.~Debbah, ``Holographic {MIMO} surfaces for {6G} wireless
  networks: {Opportunities}, challenges, and trends,'' {\em {IEEE} Wirel.
  Commun.}, vol.~27, no.~5, pp.~118--125, 2020.

\bibitem{9952197}
J.~An, C.~Xu, Q.~Wu, D.~W.~K. Ng, M.~D. Renzo, C.~Yuen, and L.~Hanzo,
  ``Codebook-based solutions for reconfigurable intelligent surfaces and their
  open challenges,'' {\em IEEE Wireless Commun.}, pp.~1--8, 2022.

\bibitem{10109654}
S.~Khisa, M.~Elhattab, C.~Assi, and S.~Sharafeddine, ``Energy consumption
  optimization in {RIS}-assisted cooperative {RSMA} cellular networks,'' {\em
  {IEEE} Trans. Commun.}, pp.~1--1, 2023.

\bibitem{DBLP:journals/twc/WuZ19}
Q.~Wu and R.~Zhang, ``Intelligent reflecting surface enhanced wireless network
  via joint active and passive beamforming,'' {\em {IEEE} Trans. Wirel.
  Commun.}, vol.~18, no.~11, pp.~5394--5409, 2019.

\bibitem{DBLP:journals/twc/ChenWCNH23}
G.~Chen, Q.~Wu, W.~Chen, D.~W.~K. Ng, and L.~Hanzo, ``{IRS}-aided wireless
  powered {MEC} systems: {TDMA} or {NOMA} for computation offloading?,'' {\em
  {IEEE} Trans. Wirel. Commun.}, vol.~22, no.~2, pp.~1201--1218, 2023.

\bibitem{10158690}
J.~An, C.~Xu, D.~W.~K. Ng, G.~C. Alexandropoulos, C.~Huang, C.~Yuen, and
  L.~Hanzo, ``Stacked intelligent metasurfaces for efficient holographic {MIMO}
  communications in {6G},'' {\em IEEE J. Sel. Areas Commun.}, pp.~1--1, 2023.

\bibitem{10088448}
M.~Misbah, Z.~Kaleem, W.~Khalid, C.~Yuen, and A.~Jamalipour, ``Phase and {3D}
  placement optimization for rate enhancement in {RIS}-assisted {UAV}
  networks,'' {\em {IEEE} Wirel. Commun. Lett.}, pp.~1--1, 2023.

\bibitem{10050148}
M.~Zeng, X.~Ning, W.~Wang, Q.~Wu, and Z.~Fei, ``{RIS} aided {NR-U} and {WiFi}
  coexistence in single cell and multiple cell networks on unlicensed bands,''
  {\em {IEEE} Trans. Green Commun. Netw.}, pp.~1--1, 2023.

\bibitem{DBLP:journals/wcl/GaoCHY21}
H.~Gao, K.~Cui, C.~Huang, and C.~Yuen, ``Robust beamforming for {RIS}-assisted
  wireless communications with discrete phase shifts,'' {\em {IEEE} Wirel.
  Commun. Lett.}, vol.~10, no.~12, pp.~2619--2623, 2021.

\bibitem{DBLP:journals/twc/NingHWWYYZ24}
Z.~Ning, H.~Hu, X.~Wang, Q.~Wu, C.~Yuen, F.~R. Yu, and Y.~Zhang, ``Joint user
  association, interference cancellation, and power control for multi-{IRS}
  assisted {UAV} communications,'' {\em {IEEE} Trans. Wirel. Commun.}, vol.~23,
  no.~10, pp.~13408--13423, 2024.

\bibitem{DBLP:journals/jsac/CaoYHYRNH21}
X.~Cao, B.~Yang, C.~Huang, C.~Yuen, M.~D. Renzo, D.~Niyato, and Z.~Han,
  ``Reconfigurable intelligent surface-assisted aerial-terrestrial
  communications via multi-task learning,'' {\em {IEEE} J. Sel. Areas Commun.},
  vol.~39, no.~10, pp.~3035--3050, 2021.

\bibitem{DBLP:conf/globecom/ZhangSB19}
Q.~Zhang, W.~Saad, and M.~Bennis, ``Reflections in the sky: {Millimeter} wave
  communication with {UAV}-carried intelligent reflectors,'' in {\em 2019
  {IEEE} Global Communications Conference, {GLOBECOM} 2019, Waikoloa, HI, USA,
  December 9-13, 2019}, pp.~1--6, {IEEE}, 2019.

\bibitem{DBLP:journals/jsac/LiuLC21}
X.~Liu, Y.~Liu, and Y.~Chen, ``Machine learning empowered trajectory and
  passive beamforming design in {UAV-RIS} wireless networks,'' {\em {IEEE} J.
  Sel. Areas Commun.}, vol.~39, no.~7, pp.~2042--2055, 2021.

\bibitem{DBLP:journals/twc/KhaliliMZJYJ22}
A.~Khalili, E.~M. Monfared, S.~Zargari, M.~R. Javan, N.~M. Yamchi, and E.~A.
  Jorswieck, ``Resource management for transmit power minimization in
  {UAV}-assisted {RIS} {HetNets} supported by dual connectivity,'' {\em {IEEE}
  Trans. Wirel. Commun.}, vol.~21, no.~3, pp.~1806--1822, 2022.

\bibitem{DBLP:journals/twc/ChengPHAYD22}
Y.~Cheng, W.~Peng, C.~Huang, G.~C. Alexandropoulos, C.~Yuen, and M.~Debbah,
  ``{RIS}-aided wireless communications: Extra degrees of freedom via rotation
  and location optimization,'' {\em {IEEE} Trans. Wirel. Commun.}, vol.~21,
  no.~8, pp.~6656--6671, 2022.

\bibitem{DBLP:journals/jsac/WuXZNASS21a}
Q.~Wu, J.~Xu, Y.~Zeng, D.~W.~K. Ng, N.~Al{-}Dhahir, R.~Schober, and A.~L.
  Swindlehurst, ``A comprehensive overview on {5G}-and-beyond networks with
  {UAVs}: {From} communications to sensing and intelligence,'' {\em {IEEE} J.
  Sel. Areas Commun.}, vol.~39, no.~10, pp.~2912--2945, 2021.

\bibitem{10793319}
M.~T. Dabiri, M.~Hasna, S.~Althunibat, K.~Qaraqe, and M.-S. Alouini, ``A
  balloon-based {UAV}-aided non-terrestrial sectorized network for post
  disaster cellular coverage: {A} dynamic environment perspective,'' in {\em
  2024 7th International Conference on Advanced Communication Technologies and
  Networking (CommNet)}, pp.~1--7, 2024.

\bibitem{DBLP:journals/icl/SudheeshMMSM18}
P.~G. Sudheesh, M.~Mozaffari, M.~Magarini, W.~Saad, and
  P.~Muthuchidambaranathan, ``Sum-rate analysis for high altitude platform
  {(HAP)} drones with tethered balloon relay,'' {\em {IEEE} Commun. Lett.},
  vol.~22, no.~6, pp.~1240--1243, 2018.

\bibitem{DBLP:journals/twc/YangCSXSPC22}
Z.~Yang, M.~Chen, W.~Saad, W.~Xu, M.~Shikh{-}Bahaei, H.~V. Poor, and S.~Cui,
  ``Energy-efficient wireless communications with distributed reconfigurable
  intelligent surfaces,'' {\em {IEEE} Trans. Wirel. Commun.}, vol.~21, no.~1,
  pp.~665--679, 2022.

\bibitem{DBLP:conf/iecon/ZhouQ018}
Y.~Zhou, G.~Qin, and F.~Lin, ``Development of nano {UAV} platform for
  navigation in {GPS}-denied environment using snapdragon,'' in {\em {IECON}
  2018 - 44th Annual Conference of the {IEEE} Industrial Electronics Society,
  Washington, DC, USA, October 21-23, 2018}, pp.~5642--5647, {IEEE}, 2018.

\bibitem{10025789}
B.~Sagir, E.~Aydin, and H.~Ilhan, ``Deep-learning assisted {IoT} based {RIS}
  for cooperative communications,'' {\em {IEEE} Internet Things J.}, pp.~1--1,
  2023.

\bibitem{DBLP:journals/tits/ZhaoSNXGZ24}
H.~Zhao, W.~Sun, Y.~Ni, W.~Xia, G.~Gui, and C.~Zhu, ``Deep deterministic policy
  gradient-based rate maximization for {RIS-UAV}-assisted vehicular
  communication networks,'' {\em {IEEE} Trans. Intell. Transp. Syst.}, vol.~25,
  no.~11, pp.~15732--15744, 2024.

\bibitem{DBLP:journals/wcl/MohamedHH22}
E.~M. Mohamed, S.~Hashima, and K.~Hatano, ``Energy aware multiarmed bandit for
  millimeter wave-based {UAV} mounted {RIS} networks,'' {\em {IEEE} Wirel.
  Commun. Lett.}, vol.~11, no.~6, pp.~1293--1297, 2022.

\bibitem{DBLP:conf/wcnc/WuGLGKK24}
M.~Wu, K.~Guo, Z.~Lin, S.~Garg, K.~Kaur, and G.~Kaddoum, ``Energy efficiency
  optimization in {RIS}-assisted {ISATRNs} with {RSMA:} {A} federated deep
  reinforcement learning approach,'' in {\em {IEEE} Wireless Communications and
  Networking Conference, {WCNC} 2024, Dubai, United Arab Emirates, April 21-24,
  2024}, pp.~1--6, {IEEE}, 2024.

\bibitem{DBLP:conf/globecom/GeZW21}
L.~Ge, H.~Zhang, and J.~Wang, ``Joint placement and beamforming design in
  multi-{UAV-IRS} assisted multiuser communication,'' in {\em {IEEE} Global
  Communications Conference, {GLOBECOM} 2021, Madrid, Spain, December 7-11,
  2021}, pp.~1--6, {IEEE}, 2021.

\bibitem{song2024enhancing}
X.~Song, D.~Li, J.~Tang, N.~Zhao, Z.~Yang, Z.~Yin, and Z.~Wu, ``Enhancing
  cell-free network: {Joint} beamforming and location optimization via
  {UAV-IRS},'' {\em {IEEE} Trans. Veh. Technol.}, 2024.

\bibitem{DBLP:journals/sj/AdamOWMAML24}
A.~B.~M. Adam, M.~A. Ouamri, X.~Wan, M.~S.~A. Muthanna, R.~Alkanhel,
  A.~Muthanna, and X.~Li, ``Secure communication in {UAV-RIS}-empowered
  multiuser networks: {Joint} beamforming, phase shift, and {UAV} trajectory
  optimization,'' {\em {IEEE} Syst. J.}, vol.~18, no.~2, pp.~1009--1019, 2024.

\bibitem{DBLP:journals/wcl/WangNTEN23}
W.~Wang, W.~Ni, H.~Tian, Y.~C. Eldar, and D.~Niyato, ``{UAV}-mounted
  multi-functional {RIS} for combating eavesdropping in wireless networks,''
  {\em {IEEE} Wirel. Commun. Lett.}, vol.~12, no.~10, pp.~1667--1671, 2023.

\bibitem{DBLP:journals/icl/0002TTD0HK23}
Y.~Xiao, D.~Tyrovolas, S.~A. Tegos, P.~D. Diamantoulakis, Z.~Ma, L.~Hao, and
  G.~K. Karagiannidis, ``Solar powered {UAV}-mounted {RIS} networks,'' {\em
  {IEEE} Commun. Lett.}, vol.~27, no.~6, pp.~1565--1569, 2023.

\bibitem{DBLP:journals/tvt/LiangZDSLW24}
X.~Liang, Z.~Zhang, Q.~Deng, F.~Shu, S.~Liu, and J.~Wang, ``Joint trajectory
  and primary-secondary transmission design for {UAV}-carried-{IRS} assisted
  underlay {CR} networks,'' {\em {IEEE} Trans. Veh. Technol.}, vol.~73, no.~11,
  pp.~17848--17853, 2024.

\bibitem{DBLP:journals/iotj/TyrovolasMBMTDILK24}
D.~Tyrovolas, N.~A. Mitsiou, T.~G. Boufikos, P.~Mekikis, S.~A. Tegos, P.~D.
  Diamantoulakis, S.~Ioannidis, C.~K. Liaskos, and G.~K. Karagiannidis,
  ``Energy-aware trajectory optimization for {UAV}-mounted {RIS} and
  full-duplex relay,'' {\em {IEEE} Internet Things J.}, vol.~11, no.~13,
  pp.~24259--24272, 2024.

\bibitem{DBLP:journals/twc/AbouamerM22}
M.~S. Abouamer and P.~Mitran, ``Joint uplink-downlink resource allocation for
  multiuser {IRS}-assisted systems,'' {\em {IEEE} Trans. Wirel. Commun.},
  vol.~21, no.~12, pp.~10918--10933, 2022.

\bibitem{DBLP:journals/ojcs/ShehabCKAT22}
M.~J. Shehab, B.~S. Ciftler, T.~Khattab, M.~M. Abdallah, and D.~Trinchero,
  ``Deep reinforcement learning powered {IRS}-assisted downlink {NOMA},'' {\em
  {IEEE} Open J. Commun. Soc.}, vol.~3, pp.~729--739, 2022.

\bibitem{DBLP:journals/wcl/WangLSFS20}
H.~Wang, C.~Liu, Z.~Shi, Y.~Fu, and R.~Song, ``On power minimization for
  {IRS}-aided downlink {NOMA} systems,'' {\em {IEEE} Wirel. Commun. Lett.},
  vol.~9, no.~11, pp.~1808--1811, 2020.

\bibitem{DBLP:conf/vtc/Nasir24}
A.~A. Nasir, ``Secure and energy-efficient mobile edge computing with
  {UAV}-mounted- {RIS} assistance,'' in {\em 99th {IEEE} Vehicular Technology
  Conference, {VTC} Spring 2024, Singapore, June 24-27, 2024}, pp.~1--5,
  {IEEE}, 2024.

\bibitem{DBLP:journals/wcl/ZhaiDDWY22}
Z.~Zhai, X.~Dai, B.~Duo, X.~Wang, and X.~Yuan, ``Energy-efficient {UAV-}mounted
  {RIS} assisted mobile edge computing,'' {\em {IEEE} Wirel. Commun. Lett.},
  vol.~11, no.~12, pp.~2507--2511, 2022.

\bibitem{DBLP:conf/icc/MagboolKF23}
A.~Magbool, V.~Kumar, and M.~F. Flanagan, ``On energy efficiency and fairness
  maximization in {RIS}-assisted {MU-MISO} mmwave communications,'' in {\em
  {IEEE} International Conference on Communications, {ICC} 2023, Rome, Italy,
  May 28 - June 1, 2023}, pp.~5364--5369, {IEEE}, 2023.

\bibitem{DBLP:journals/twc/MaFZGY22}
X.~Ma, Y.~Fang, H.~Zhang, S.~Guo, and D.~Yuan, ``Cooperative beamforming design
  for multiple {RIS}-assisted communication systems,'' {\em {IEEE} Trans.
  Wirel. Commun.}, vol.~21, no.~12, pp.~10949--10963, 2022.

\bibitem{DBLP:journals/jsac/LiuLCP21}
X.~Liu, Y.~Liu, Y.~Chen, and H.~V. Poor, ``{RIS} enhanced massive
  non-orthogonal multiple access networks: {Deployment} and passive beamforming
  design,'' {\em {IEEE} J. Sel. Areas Commun.}, vol.~39, no.~4, pp.~1057--1071,
  2021.

\bibitem{DBLP:journals/iotj/NaeemQC23}
F.~Naeem, M.~K. Qaraqe, and H.~Celebi, ``Joint deployment design and phase
  shift of {IRS}-assisted {6G} networks: An experience-driven approach,'' {\em
  {IEEE} Internet Things J.}, vol.~10, no.~20, pp.~17647--17655, 2023.

\bibitem{DBLP:journals/tvt/LinYZXHN24}
K.~Lin, H.~Yang, M.~Zheng, L.~Xiao, C.~Huang, and D.~Niyato, ``Penalized
  reinforcement learning-based energy-efficient {UAV-RIS} assisted maritime
  uplink communications against jamming,'' {\em {IEEE} Trans. Veh. Technol.},
  vol.~73, no.~10, pp.~15768--15773, 2024.

\bibitem{DBLP:conf/ictc/ZhaoMLP22}
K.~Zhao, H.~Mei, S.~Lyu, and L.~Peng, ``Joint optimization of multiple
  {UAV}-mounted {RISs} deployment and {RIS} elements allocation,'' in {\em 13th
  International Conference on Information and Communication Technology
  Convergence, {ICTC} 2022, Jeju Island, Korea, Republic of, October 19-21,
  2022}, pp.~1193--1197, {IEEE}, 2022.

\bibitem{DBLP:journals/twc/FengXYX22}
T.~Feng, L.~Xie, J.~Yao, and J.~Xu, ``{UAV}-enabled data collection for
  wireless sensor networks with distributed beamforming,'' {\em {IEEE} Trans.
  Wirel. Commun.}, vol.~21, no.~2, pp.~1347--1361, 2022.

\bibitem{DBLP:journals/jstsp/ZhaoXYWS22}
Y.~Zhao, W.~Xu, X.~You, N.~Wang, and H.~Sun, ``Cooperative reflection and
  synchronization design for distributed multiple-{RIS} communications,'' {\em
  {IEEE} J. Sel. Top. Signal Process.}, vol.~16, no.~5, pp.~980--994, 2022.

\bibitem{DBLP:journals/icl/FaisalADN22}
A.~Faisal, I.~Al{-}Nahhal, O.~A. Dobre, and T.~M.~N. Ngatched, ``Deep
  reinforcement learning for {RIS}-assisted {FD} systems: {Single} or
  distributed {RIS}?,'' {\em {IEEE} Commun. Lett.}, vol.~26, no.~7,
  pp.~1563--1567, 2022.

\bibitem{DBLP:journals/twc/HuangZADY19}
C.~Huang, A.~Zappone, G.~C. Alexandropoulos, M.~Debbah, and C.~Yuen,
  ``Reconfigurable intelligent surfaces for energy efficiency in wireless
  communication,'' {\em {IEEE} Trans. Wirel. Commun.}, vol.~18, no.~8,
  pp.~4157--4170, 2019.

\bibitem{DBLP:journals/tcom/WeiHAYZD21}
L.~Wei, C.~Huang, G.~C. Alexandropoulos, C.~Yuen, Z.~Zhang, and M.~Debbah,
  ``Channel estimation for {RIS}-empowered multi-user {MISO} wireless
  communications,'' {\em {IEEE} Trans. Commun.}, vol.~69, no.~6,
  pp.~4144--4157, 2021.

\bibitem{DBLP:journals/icl/VuK23}
T.~Vu and S.~Kim, ``Performance analysis of full-duplex two-way {RIS}-based
  systems with imperfect {CSI} and discrete phase-shift design,'' {\em {IEEE}
  Commun. Lett.}, vol.~27, no.~2, pp.~512--516, 2023.

\bibitem{DBLP:journals/tcom/AnXGH22}
J.~An, C.~Xu, L.~Gan, and L.~Hanzo, ``Low-complexity channel estimation and
  passive beamforming for {RIS}-assisted {MIMO} systems relying on discrete
  phase shifts,'' {\em {IEEE} Trans. Commun.}, vol.~70, no.~2, pp.~1245--1260,
  2022.

\bibitem{DBLP:journals/jsac/ChenZBZA21}
S.~Chen, J.~Zhang, E.~Bj{\"{o}}rnson, J.~Zhang, and B.~Ai, ``Structured massive
  access for scalable cell-free massive {MIMO} systems,'' {\em {IEEE} J. Sel.
  Areas Commun.}, vol.~39, no.~4, pp.~1086--1100, 2021.

\bibitem{DBLP:journals/wcl/SongZWYT22}
X.~Song, Y.~Zhao, Z.~Wu, Z.~Yang, and J.~Tang, ``Joint trajectory and
  communication design for {IRS}-assisted {UAV} networks,'' {\em {IEEE} Wirel.
  Commun. Lett.}, vol.~11, no.~7, pp.~1538--1542, 2022.

\bibitem{DBLP:journals/tcom/Al-JarrahAAIA21}
M.~A. Al{-}Jarrah, A.~Al{-}Dweik, E.~Alsusa, Y.~Iraqi, and M.~Alouini, ``On the
  performance of {IRS}-assisted multi-layer {UAV} communications with imperfect
  phase compensation,'' {\em {IEEE} Trans. Commun.}, vol.~69, no.~12,
  pp.~8551--8568, 2021.

\bibitem{DBLP:journals/pieee/ZengWZ19}
Y.~Zeng, Q.~Wu, and R.~Zhang, ``Accessing from the sky: {A} tutorial on {UAV}
  communications for 5{G} and beyond,'' {\em Proc. {IEEE}}, vol.~107, no.~12,
  pp.~2327--2375, 2019.

\bibitem{DBLP:journals/twc/ZengXZ19}
Y.~Zeng, J.~Xu, and R.~Zhang, ``Energy minimization for wireless communication
  with rotary-wing {UAV},'' {\em {IEEE} Trans. Wirel. Commun.}, vol.~18, no.~4,
  pp.~2329--2345, 2019.

\bibitem{DBLP:journals/tcom/PanLSFLY23}
H.~Pan, Y.~Liu, G.~Sun, J.~Fan, S.~Liang, and C.~Yuen, ``Joint power and {3D}
  trajectory optimization for {UAV}-enabled wireless powered communication
  networks with obstacles,'' {\em {IEEE} Trans. Commun.}, vol.~71, no.~4,
  pp.~2364--2380, 2023.

\bibitem{DBLP:journals/iotj/FengWHY24}
Z.~Feng, D.~Wu, M.~Huang, and C.~Yuen, ``Graph-attention-based reinforcement
  learning for trajectory design and resource assignment in
  multi-{UAV}-assisted communication,'' {\em {IEEE} Internet Things J.},
  vol.~11, no.~16, pp.~27421--27434, 2024.

\bibitem{DBLP:journals/tec/DebAPM02}
K.~Deb, S.~Agrawal, A.~Pratap, and T.~Meyarivan, ``A fast and elitist
  multiobjective genetic algorithm: {NSGA-II},'' {\em {IEEE} Trans. Evol.
  Comput.}, vol.~6, no.~2, pp.~182--197, 2002.

\bibitem{zhang2024uav}
C.~Zhang, G.~Sun, Q.~Wu, J.~Li, S.~Liang, D.~Niyato, and V.~C. Leung, ``{UAV}
  swarm-enabled collaborative secure relay communications with time-domain
  colluding eavesdropper,'' {\em {IEEE} Trans. Mob. Comput.}, 2024.

\bibitem{DBLP:journals/tec/ZhangTCJ15}
X.~Zhang, Y.~Tian, R.~Cheng, and Y.~Jin, ``An efficient approach to
  nondominated sorting for evolutionary multiobjective optimization,'' {\em
  {IEEE} Trans. Evol. Comput.}, vol.~19, no.~2, pp.~201--213, 2015.

\bibitem{DBLP:journals/tec/YueQL18}
C.~Yue, B.~Qu, and J.~Liang, ``A multiobjective particle swarm optimizer using
  ring topology for solving multimodal multiobjective problems,'' {\em {IEEE}
  Trans. Evol. Comput.}, vol.~22, no.~5, pp.~805--817, 2018.

\bibitem{DBLP:journals/tcom/ZhiPRW22}
K.~Zhi, C.~Pan, H.~Ren, and K.~Wang, ``Power scaling law analysis and phase
  shift optimization of {RIS}-aided massive {MIMO} systems with statistical
  {CSI},'' {\em {IEEE} Trans. Commun.}, vol.~70, no.~5, pp.~3558--3574, 2022.

\bibitem{DBLP:journals/tnse/LongZDPOX23}
S.~Long, Y.~Zhang, Q.~Deng, T.~Pei, J.~Ouyang, and Z.~Xia, ``An efficient task
  offloading approach based on multi-objective evolutionary algorithm in
  cloud-edge collaborative environment,'' {\em {IEEE} Trans. Netw. Sci. Eng.},
  vol.~10, no.~2, pp.~645--657, 2023.

\bibitem{DBLP:journals/tec/CoelloPL04}
C.~A.~C. Coello, G.~T. Pulido, and M.~S. Lechuga, ``Handling multiple
  objectives with particle swarm optimization,'' {\em {IEEE} Trans. Evol.
  Comput.}, vol.~8, no.~3, pp.~256--279, 2004.

\bibitem{DBLP:journals/eswa/MirjaliliSMC16}
S.~Mirjalili, S.~Saremi, S.~M. Mirjalili, and L.~dos Santos~Coelho,
  ``Multi-objective grey wolf optimizer: {A} novel algorithm for
  multi-criterion optimization,'' {\em Expert Syst. Appl.}, vol.~47,
  pp.~106--119, 2016.

\bibitem{DBLP:journals/tec/Jensen03a}
M.~T. Jensen, ``Reducing the run-time complexity of multiobjective eas: The
  {NSGA-II} and other algorithms,'' {\em {IEEE} Trans. Evol. Comput.}, vol.~7,
  no.~5, pp.~503--515, 2003.

\bibitem{DBLP:journals/tcom/WuZ20}
Q.~Wu and R.~Zhang, ``Beamforming optimization for wireless network aided by
  intelligent reflecting surface with discrete phase shifts,'' {\em {IEEE}
  Trans. Commun.}, vol.~68, no.~3, pp.~1838--1851, 2020.

\bibitem{DBLP:journals/tcyb/ZhangGLSZT22}
C.~Zhang, L.~Gao, X.~Li, W.~Shen, J.~Zhou, and K.~C. Tan, ``Resetting weight
  vectors in {MOEA/D} for multiobjective optimization problems with
  discontinuous pareto front,'' {\em {IEEE} Trans. Cybern.}, vol.~52, no.~9,
  pp.~9770--9783, 2022.

\bibitem{DBLP:journals/tpds/ZhouRXF23}
Y.~Zhou, Y.~Ren, M.~Xu, and G.~Feng, ``An improved {NSGA-III} algorithm based
  on deep {Q}-networks for cloud storage optimization of blockchain,'' {\em
  {IEEE} Trans. Parallel Distributed Syst.}, vol.~34, no.~5, pp.~1406--1419,
  2023.

\bibitem{DBLP:journals/twc/MehariPCDVPJMDM16}
M.~T. Mehari, E.~D. Poorter, I.~Couckuyt, D.~Deschrijver, G.~Vermeeren,
  D.~Plets, W.~Joseph, L.~Martens, T.~Dhaene, and I.~Moerman, ``Efficient
  identification of a multi-objective pareto front on a wireless
  experimentation facility,'' {\em {IEEE} Trans. Wirel. Commun.}, vol.~15,
  no.~10, pp.~6662--6675, 2016.

\bibitem{DBLP:journals/joi/Antonoyiannakis18}
M.~Antonoyiannakis, ``Impact factors and the central limit theorem: Why
  citation averages are scale dependent,'' {\em J. Informetrics}, vol.~12,
  no.~4, pp.~1072--1088, 2018.

\bibitem{10012331}
J.~Li, G.~Sun, H.~Kang, A.~Wang, S.~Liang, Y.~Liu, and Y.~Zhang,
  ``Multi-objective optimization approaches for physical layer secure
  communications based on collaborative beamforming in {UAV} networks,'' {\em
  {IEEE/ACM} Trans. Netw.}, vol.~31, no.~4, pp.~1902--1917, 2023.

\bibitem{DBLP:journals/wcl/ZhouHJMD21}
H.~Zhou, F.~Hu, M.~Juras, A.~B. Mehta, and Y.~Deng, ``Real-time video streaming
  and control of cellular-connected {UAV} system: Prototype and performance
  evaluation,'' {\em {IEEE} Wirel. Commun. Lett.}, vol.~10, no.~8,
  pp.~1657--1661, 2021.

\bibitem{jeong2005efficient}
S.~Jeong, M.~Murayama, and K.~Yamamoto, ``Efficient optimization design method
  using kriging model,'' {\em J Aircr}, vol.~42, no.~2, pp.~413--420, 2005.

\bibitem{10278101}
H.~Pan, Y.~Liu, G.~Sun, P.~Wang, and C.~Yuen, ``Resource scheduling for
  {UAVs}-aided {D2D} networks: {A} multi-objective optimization approach,''
  {\em {IEEE} Trans. Wirel. Commun.}, pp.~1--1, 2023.

\bibitem{DBLP:journals/tits/MuntahaHJH21}
S.~T. Muntaha, S.~A. Hassan, H.~Jung, and M.~S. Hossain, ``Energy efficiency
  and hover time optimization in {UAV}-based {HetNets},'' {\em {IEEE} Trans.
  Intell. Transp. Syst.}, vol.~22, no.~8, pp.~5103--5111, 2021.

\end{thebibliography}

\end{document}